\documentclass{JINST}

\title{A method for statistical comparison of histograms}

\author{S. Bityukov$^a$\thanks{Corresponding
author.}, N. Krasnikov$^b$, A. Nikitenko$^c$  and V. Smirnova$^a$\\
\llap{$^a$}Institute for high energy physics,\\ 
  142281 Protvino, Russia\\
\llap{$^b$}Institute for nuclear research RAS,\\
  117312 Moscow, Russia\\
\llap{$^c$}Imperial College,\\
  London, United Kingdom,  on leave from ITEP, Moscow, Russia\\
  E-mail: \email{Serguei.Bitioukov@cern.ch}}

\abstract{We propose an approach for testing the hypothesis that two 
realizations of the random variables in the form of histograms are taken 
from the same statistical population (i.e. that two histograms are drawn from the 
same distribution). The approach is based on the notion ``significance of deviation''.  
Our approach allows also to estimate the statistical difference between two 
histograms.}


\begin{document}

\section{Introduction}\label{sec:introduce} 

The problem  of the testing the hypothesis that two  
histograms are drawn from the same distribution  
is a very important problem in many scientific researches. For example, this problem 
 exists for  the   
monitoring of the experimental equipment in an  experiment. Several 
 approaches to formalize and resolve this problem were considered~\cite{Porter}. 
Recently, the comparison of weighted histograms was developed in paper~\cite{Gagun}. 

In this note we propose a method which allows to estimate the value 
of statistical difference between histograms.

\section{Significance of deviations}\label{sec:significance}

In paper~\cite{ACAT08} several types of significances of deviation (or significance of an 
enhancement~\cite{Frod}) between two values were considered: 

\begin{itemize}
\item[A.] expected significance of deviation between two expected realizations of random 
variables (for example, $S_{c12}$~\cite{ACAT08});

\item[B.] significance of deviation between the observed value and expected realization of 
random variable (for example, $S_{cP}$~\cite{ACAT08}); 

\item[C.] significance of deviation between two observed values. 

\end{itemize}

As shown (in particular, in paper~\cite{ACAT08}), many of these significances obey 
the distribution close to 
the standard normal distribution  if both values are 
taken from the same statistical population.    
This property is used  here for the estimation of statistical difference between two histograms. 
We consider the significance of type C in this note.

\section{Model}\label{sec:model}

Let us consider a simple model with two histograms  where 
the random variable in each bin obeys the normal distribution  

\begin{equation}
\displaystyle 
\varphi(x|n_{ik})= \frac{1}{\sqrt{2\pi} \sigma_{ik}}~e^{-\frac{(x-n_{ik})^2}{2 \sigma^2_{ik}}}\,.
\label{eq:1}
\end{equation} 

\noindent
Here the expected value in the bin $i$ is equal to $n_{ik}$ and the variance $\sigma^2_{n_{ik}}$ 
is also equal to  $n_{ik}$. 
$k$ is the histogram number ($k=1,2$).

We define the significance  as  

\begin{equation}
\hat S_i = \displaystyle 
\frac{\hat n_{i1} - \hat n_{i2}}{\sqrt{\hat \sigma^2_{n_{i1}} + \hat \sigma^2_{n_{i2}}}}. 
\label{eq:2}
\end{equation}

\noindent 
Here $\hat n_{ik}$ is an observed value in the bin $i$ of the 
histogram $k$ and  $~\hat \sigma^2_{n_{ik}}=\hat n_{ik}$. 

This model can be considered as the approximation of Poisson distribution by normal distribution. 
So, we suppose that the  values 
$\hat n_{ik},~~(i=1,2,\dots,M,~~k=1,2)$ are the  numbers of events appeared in the 
bin $i$ for the histogram $k$. 
We consider the $RMS$ (the root mean square) of the distribution 
of the significances~\footnote{We use the ROOT notation for RMS (denominator equals M) 
in order that to have possibility for comparison of histograms with one bin.} 

$$RMS = \displaystyle \sqrt{\frac{\sum_{i=1}^M{(\hat S_i - \bar S)^2}}{M}}.$$

\noindent
Here $\bar S$ is a mean value of $\hat S_i$. 
The $RMS$ has the meaning of the  ``distance measure'' 
between two histograms.
Note that the  observed value of the $RMS$ can be converted to the $p-$value. 
If total number of events $N_1$ in the histogram $1$ and 
total number of events $N_2$ in the histogram $2$ are various   
then the normalized significance is used 

\begin{equation}
\hat S_i(K) = \displaystyle 
\frac{\hat n_{i1} - K \hat n_{i2}}{\sqrt{\hat \sigma^2_{n_{i1}} + K^2 \hat \sigma^2_{n_{i2}}}}, 
\label{eq:3}
\end{equation}

\noindent
where $K = \displaystyle \frac{N_1}{N_2}$.

Let us consider several examples.  

\section{Examples}\label{sec:examples}

All calculations, Monte Carlo experiments and histograms presentation in this note are performed 
using ROOT code~\cite{ROOT}.  The number of the bins $M$ is equal to 1000. Histograms are obtained 
from independent samples. 

\subsection{Uniform distribution }

\begin{figure}[htpb]
  \begin{center}
           \resizebox{7.0cm}{!}{\includegraphics{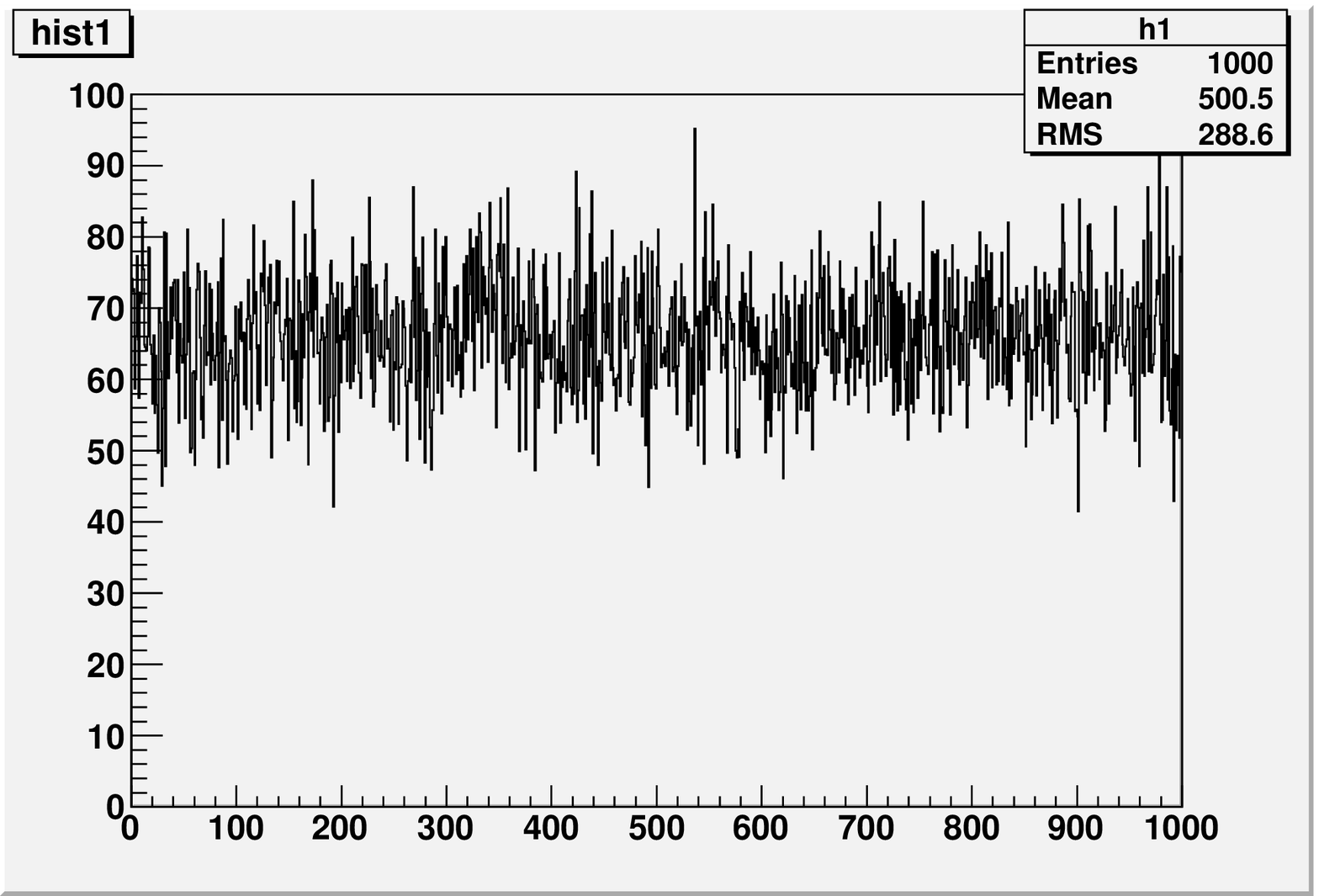}} 
           \resizebox{7.0cm}{!}{\includegraphics{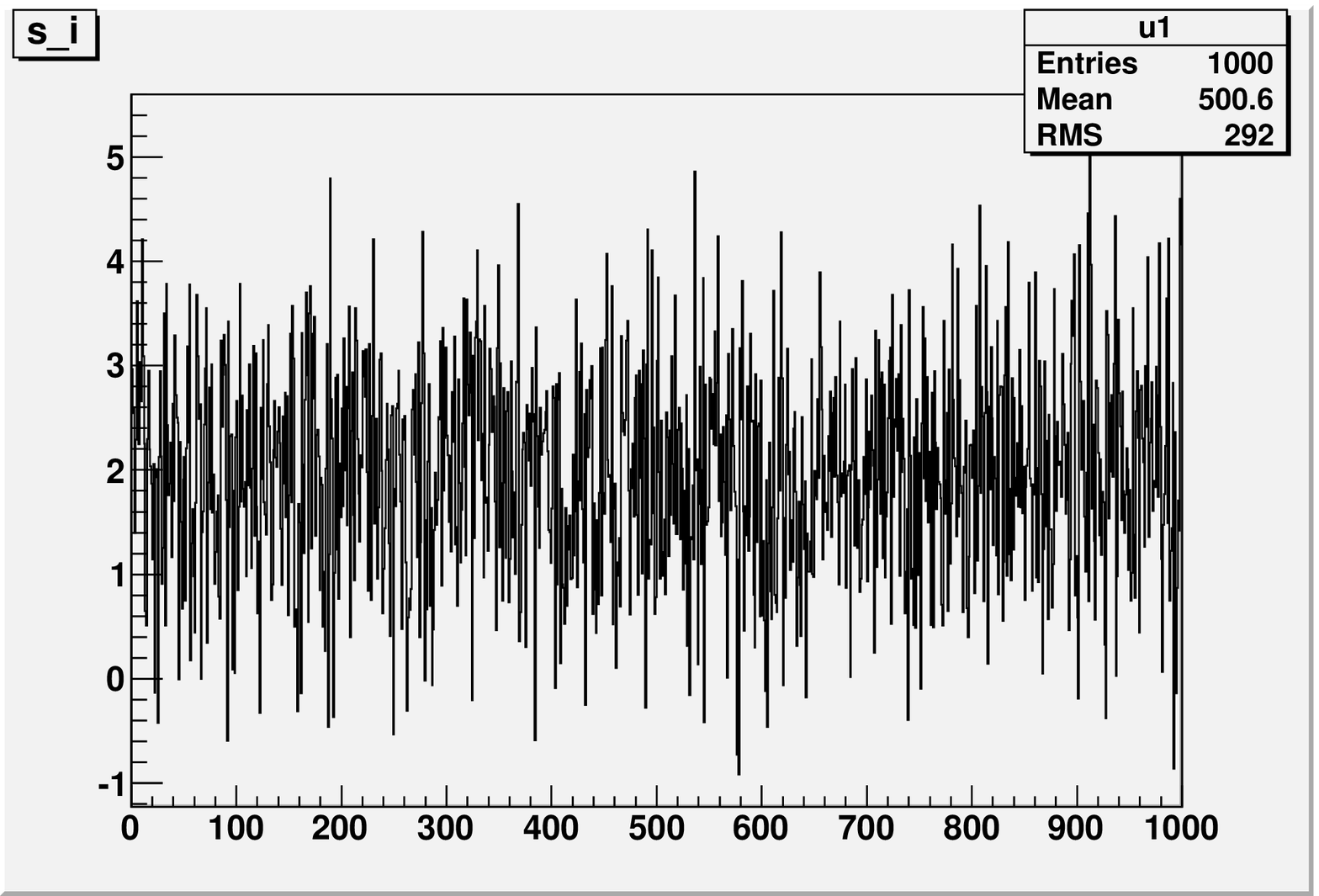}} 
           \resizebox{7.0cm}{!}{\includegraphics{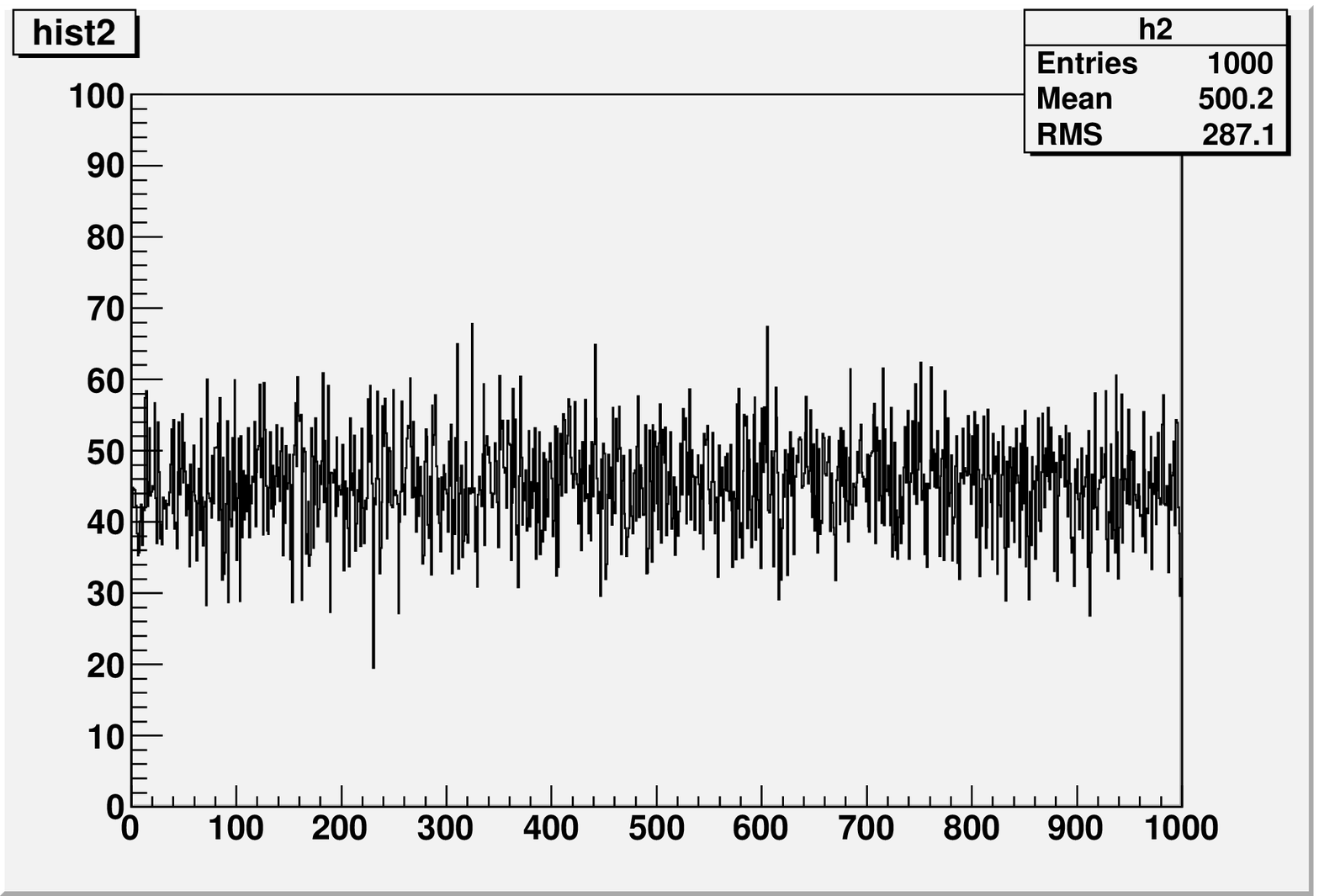}} 
           \resizebox{7.0cm}{!}{\includegraphics{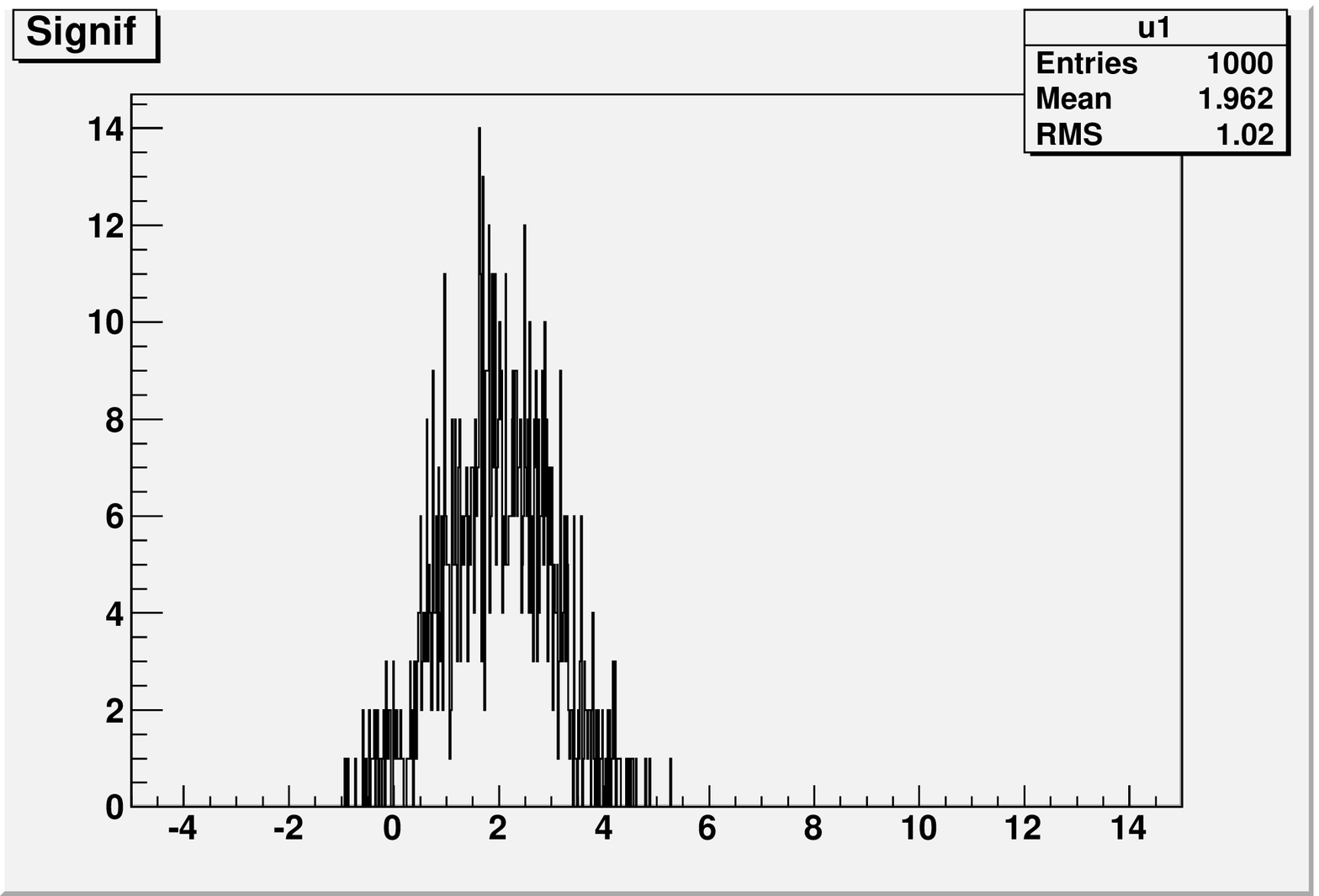}} 
\caption{Uniform distributions: the observed values $\hat n_{i1}$ in the first histogram (left,up), 
the observed values $\hat n_{i2}$ in the second histogram (left, down), 
observed significances $\hat S_{i}$ 
bin-by-bin (right, up), the distribution of observed significances (right down).}
    \label{fig:1} 
  \end{center}
\end{figure}

Consider the case when expected values $n_{i1}$ 
in the first histogram is $66$ and the expected values 
$n_{i2}$ in the second histogram is $45$ for each bin number $~i=1,2,\dots,M$. 
The results of the Monte Carlo experiment for this example are presented in Fig.~\ref{fig:1}. 

One can see that the distribution of observed significances is close to normal distribution with 
the $RMS~\sim~1$. The average significance is $\sim~1.96$, because total numbers of 
events in the  histograms are 
different. 

\begin{figure}[htpb]
  \begin{center}
           \resizebox{7.0cm}{!}{\includegraphics{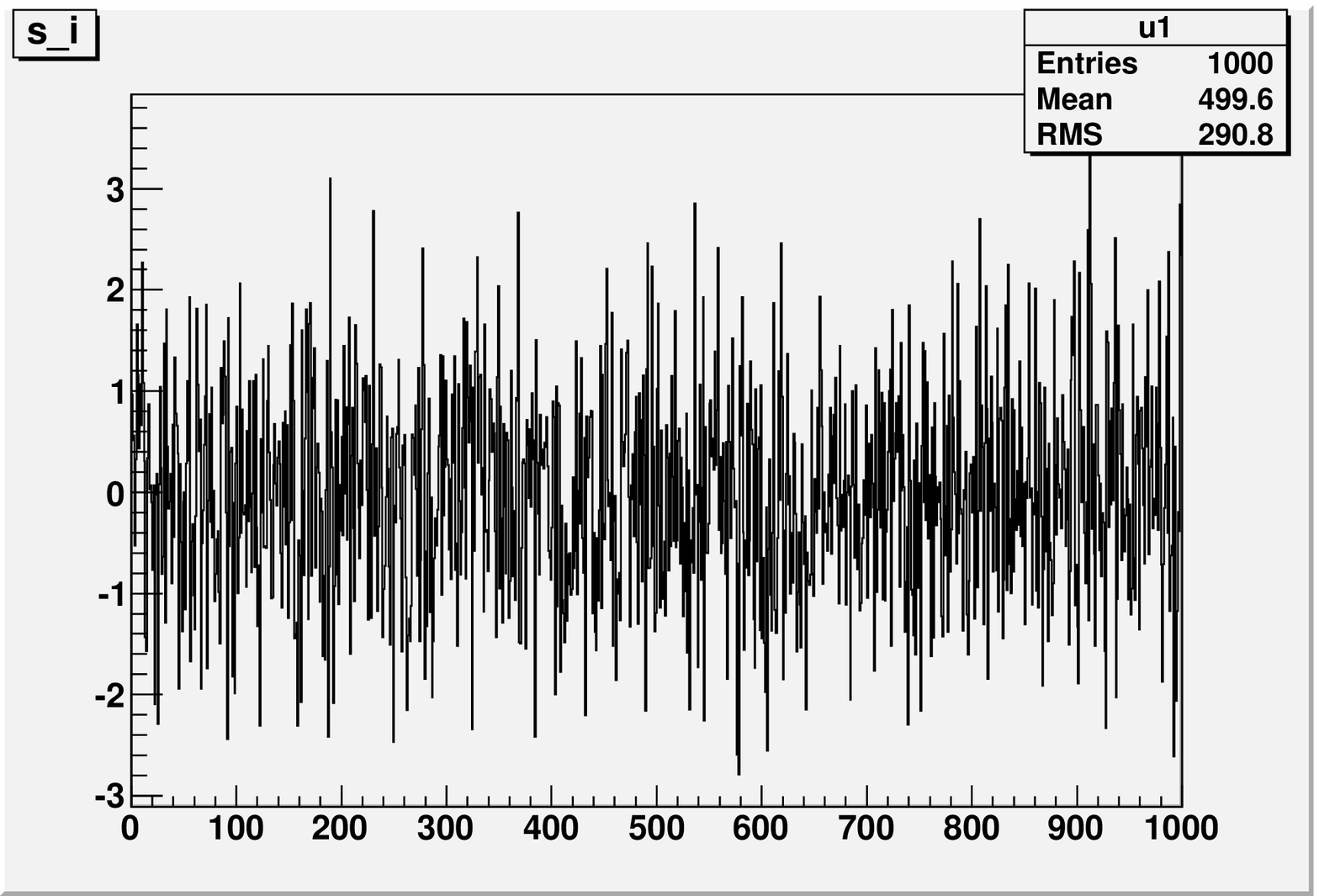}} 
           \resizebox{7.0cm}{!}{\includegraphics{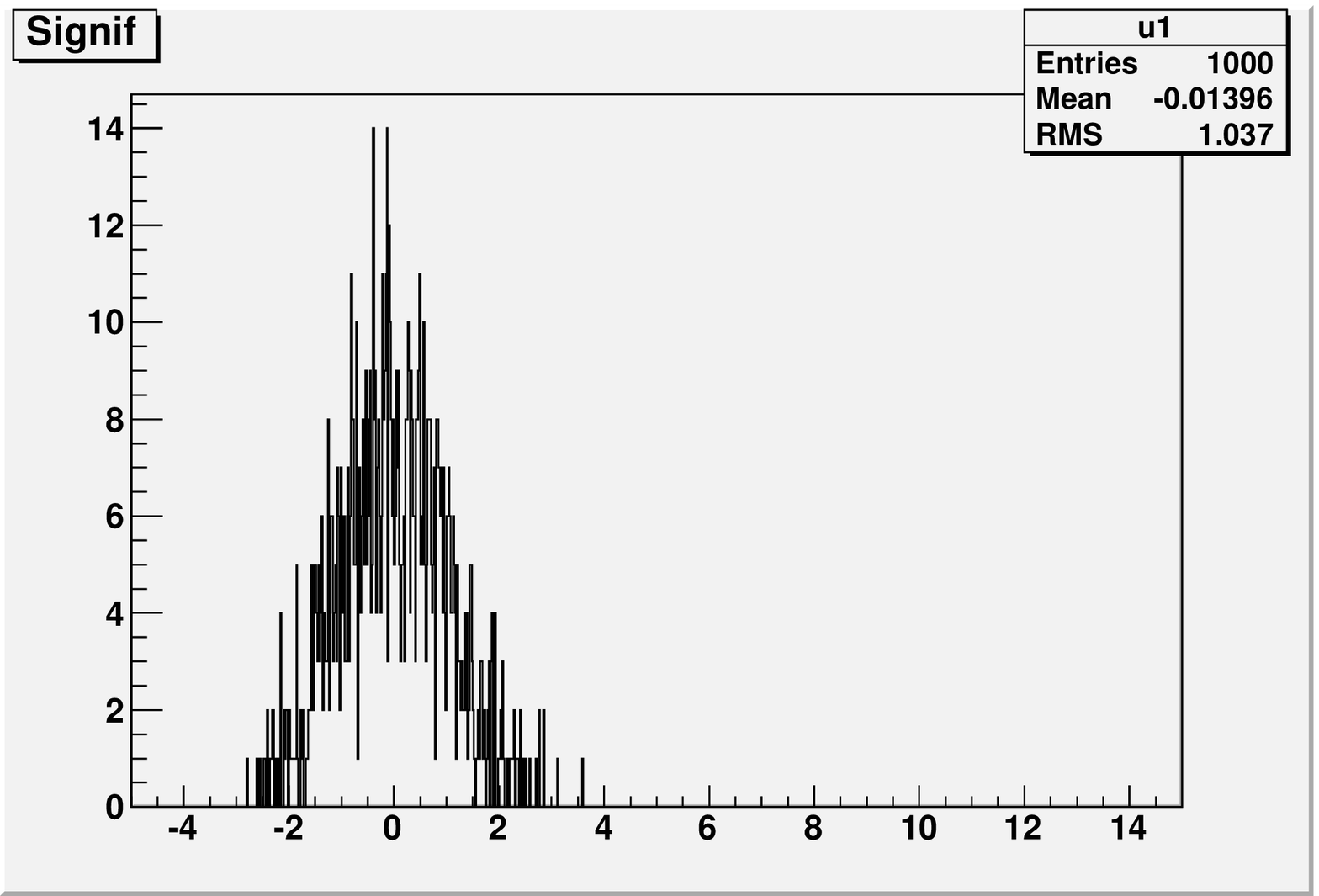}} 
\caption{Uniform distributions: observed normalized significances $\hat S_{i}$ bin-by-bin (left), 
the distribution of observed normalized significances (right).}
    \label{fig:2} 
  \end{center}
\end{figure}

In Fig.~\ref{fig:2} the corresponding histograms for normalized significances are shown.

\noindent 
The distribution of observed normalized significance is close to standard normal distribution. 

\subsection{Triangle distribution} 

\begin{figure}[htbp]
           \resizebox{7.0cm}{!}{\includegraphics{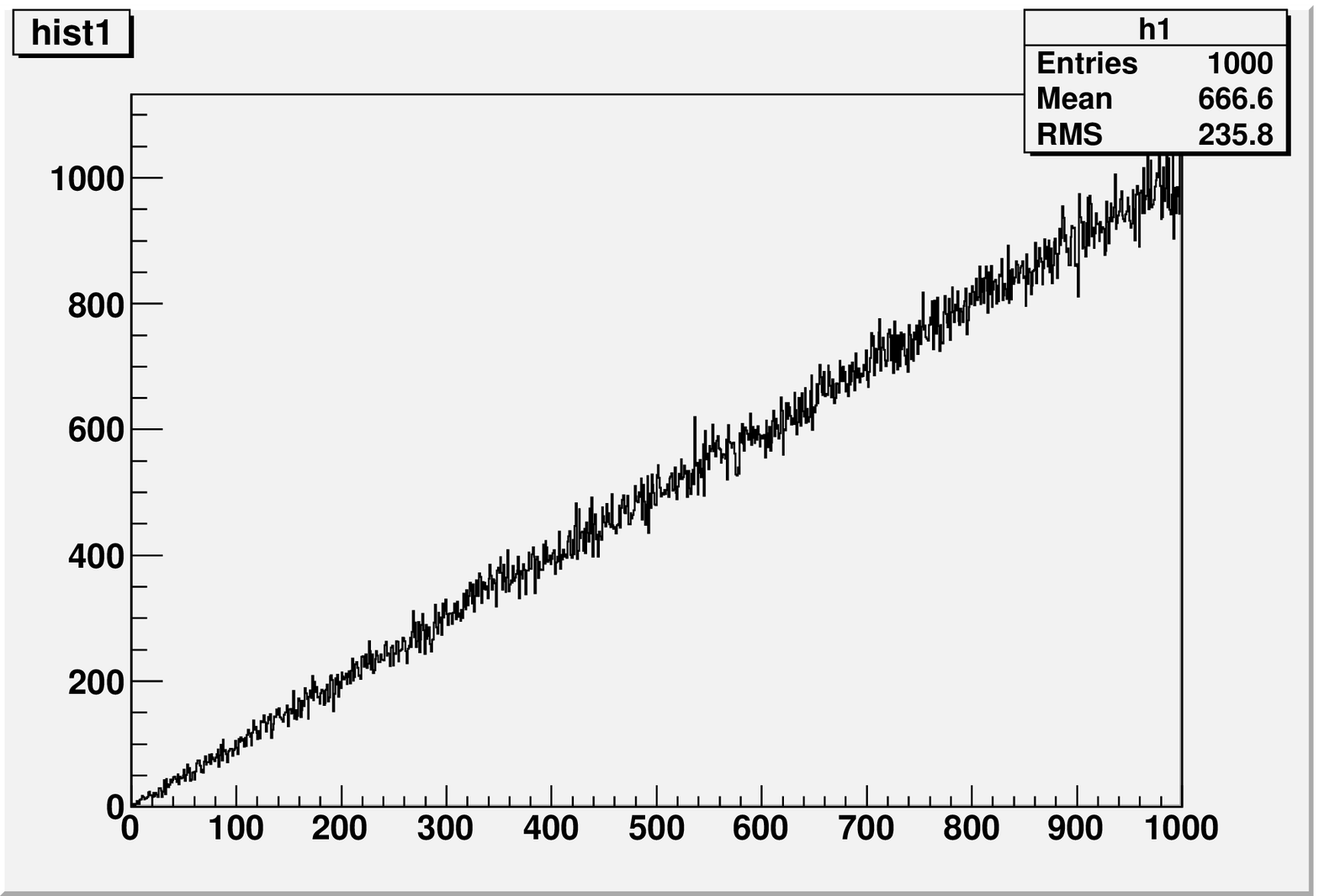}} 
           \resizebox{7.0cm}{!}{\includegraphics{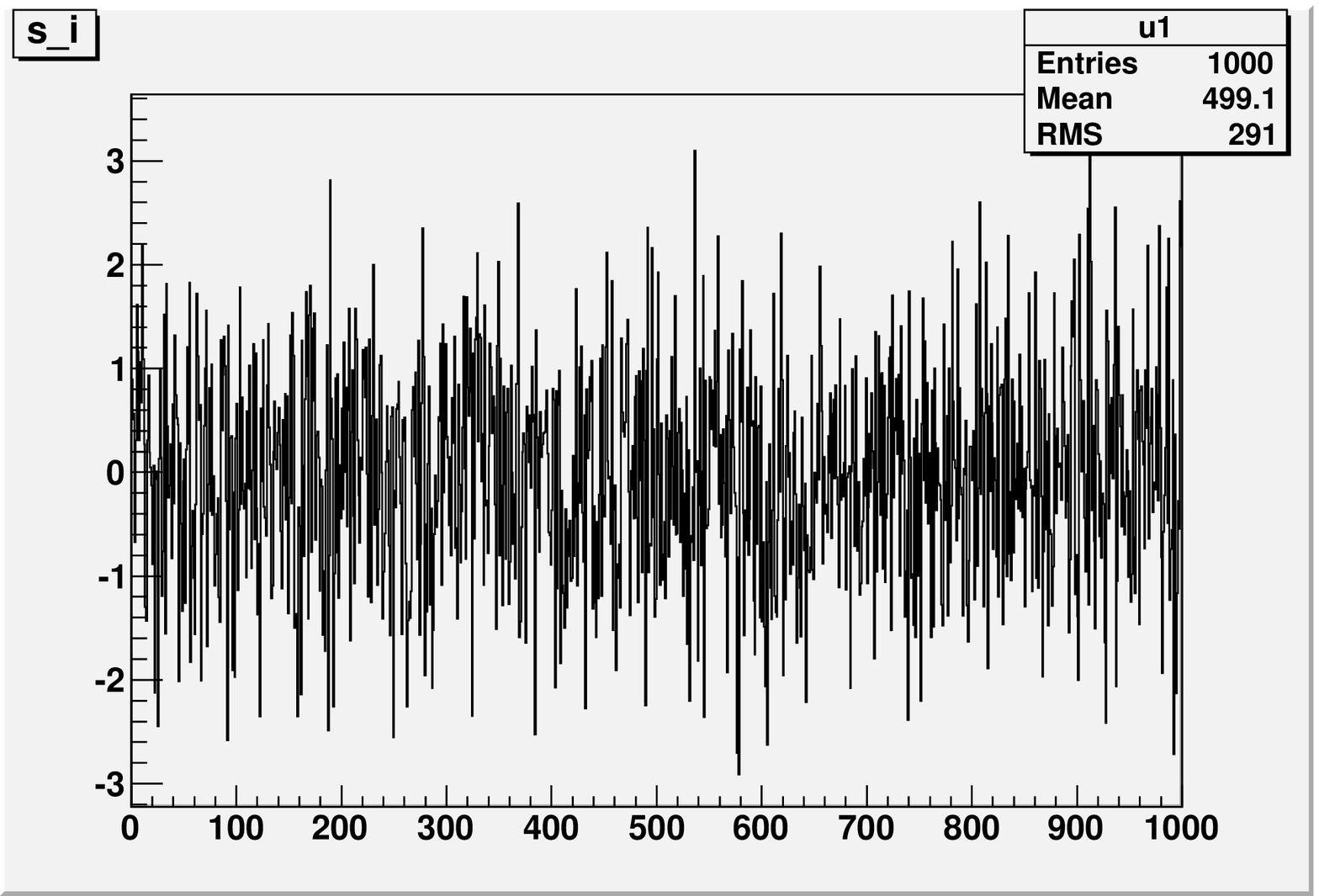}} 
           \resizebox{7.0cm}{!}{\includegraphics{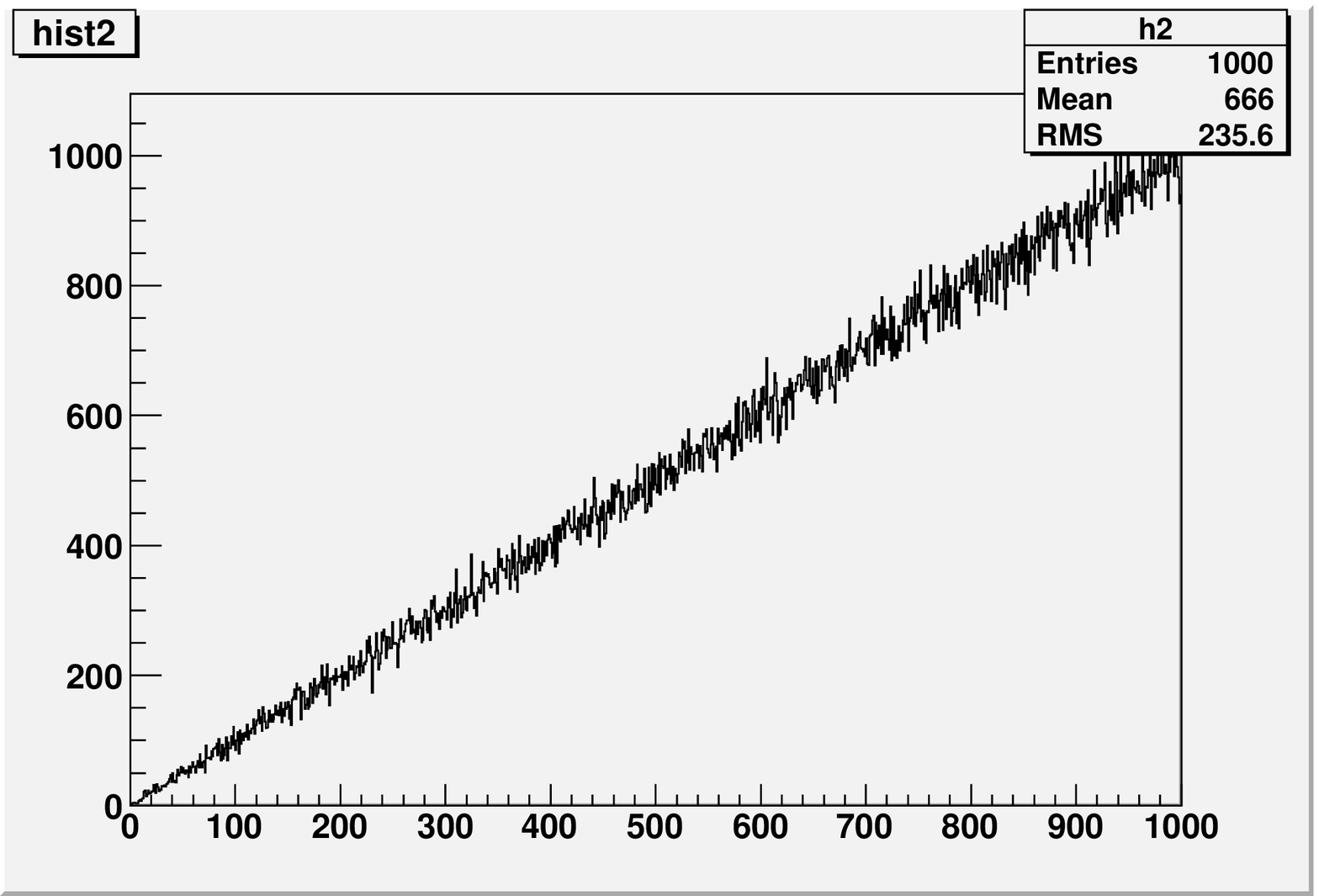}} 
           \resizebox{7.0cm}{!}{\includegraphics{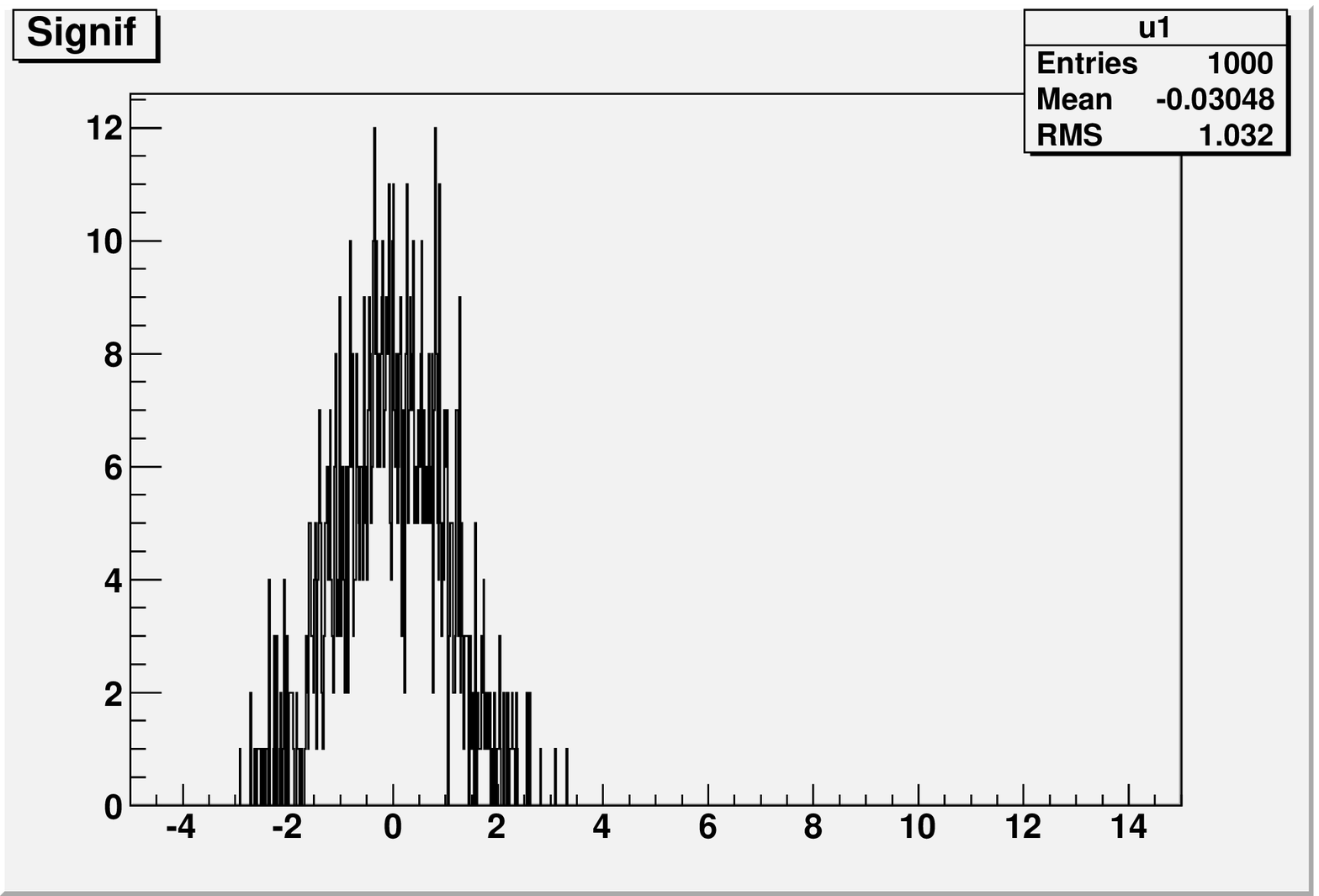}} 
\caption{Triangle distributions: the observed values $\hat n_{i1}$ in the first histogram (left,up), 
the observed values $n_{i2}$ in the second histogram (left, down), observed significances $\hat S_{i}$ 
bin-by-bin (right, up), the distribution of observed significances (right down).}
    \label{fig:3} 
\end{figure}

Consider the case when the expected values $n_{ik}$ 
in both histograms are equal to  $i$, where $i$ ($i=1,2,\dots,M$) is a bin number and  
$k$ is a histogram number ($k=1,2$). It means that the rates of events in different bins are 
different. 
One can find that in this case 
 the  distributions of observed significances is also close to standard 
normal distribution, see  Fig.~\ref{fig:3}. 
It means that the histograms which have different expected values of events 
in different bins give the distribution of significances close to standard normal distribution.

\begin{figure}[htbp]
           \resizebox{7.0cm}{!}{\includegraphics{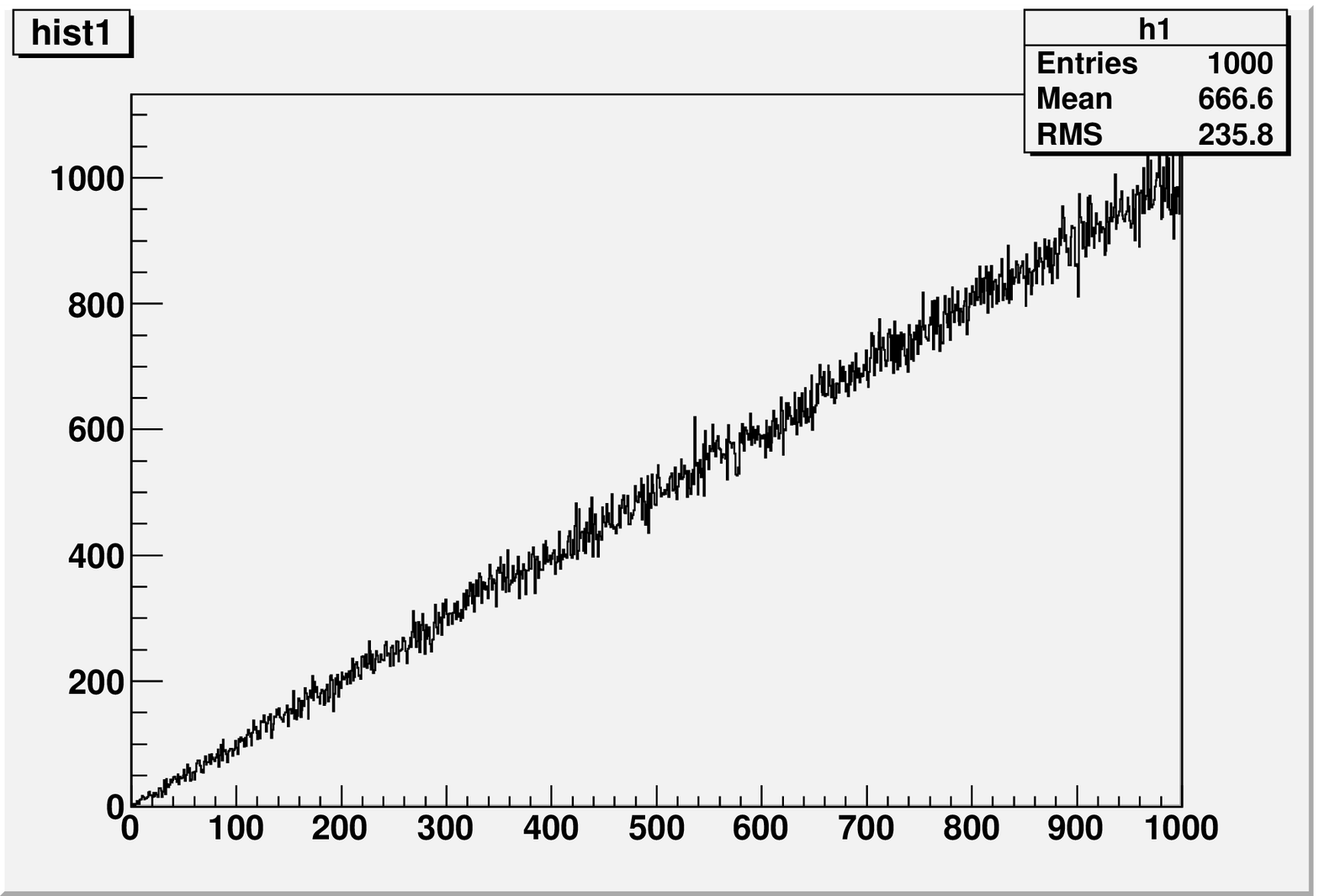}} 
           \resizebox{7.0cm}{!}{\includegraphics{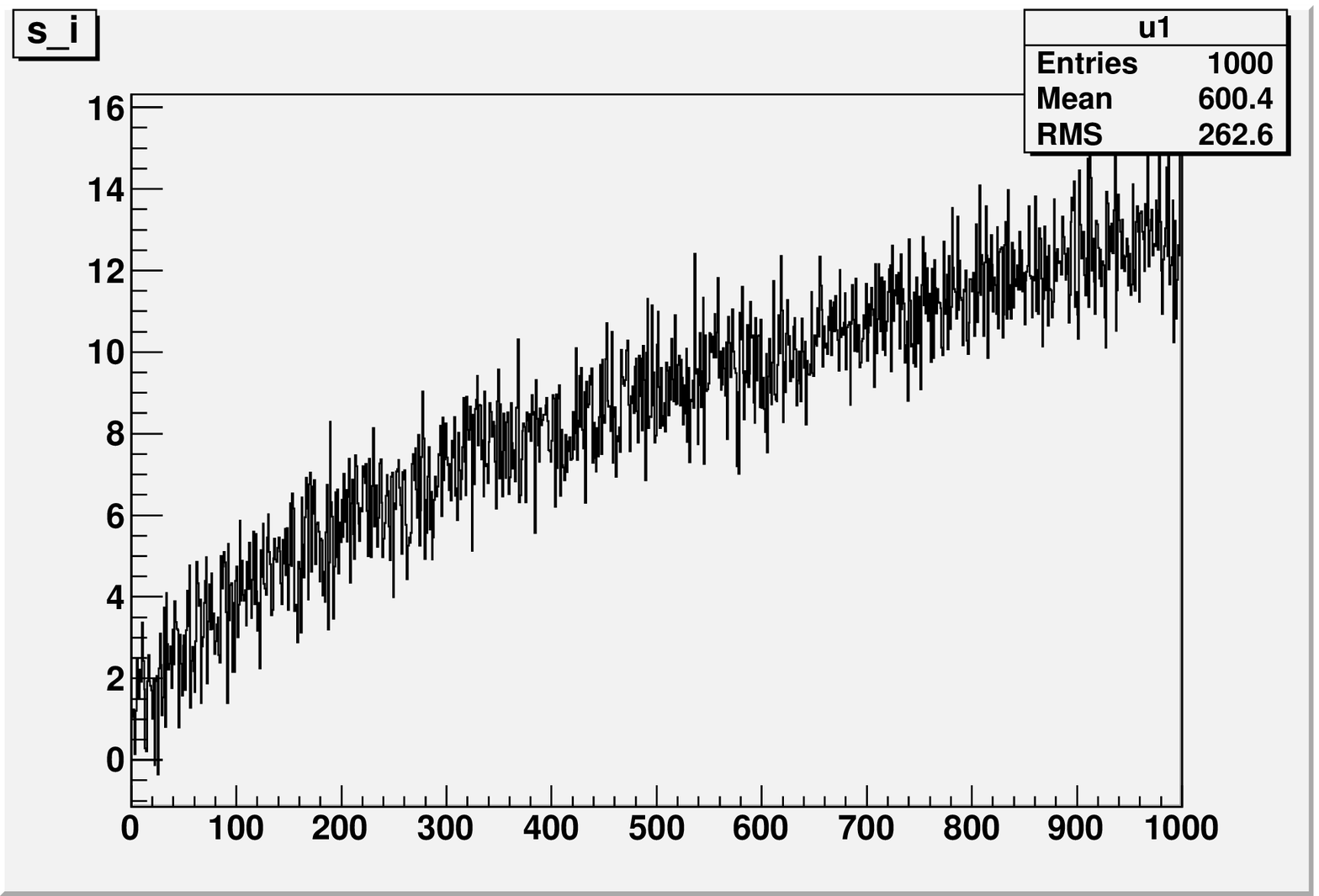}} 
           \resizebox{7.0cm}{!}{\includegraphics{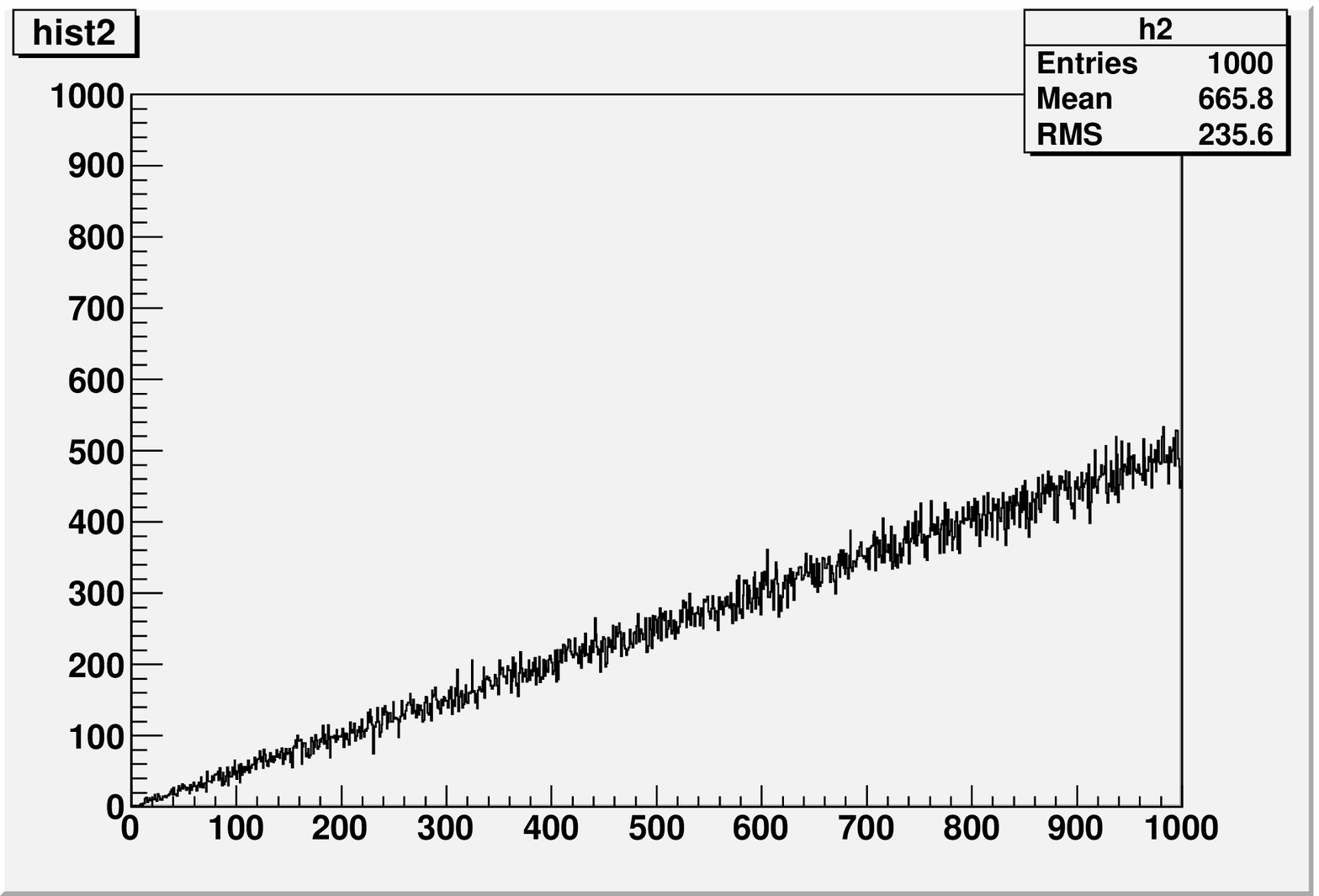}} 
           \resizebox{7.0cm}{!}{\includegraphics{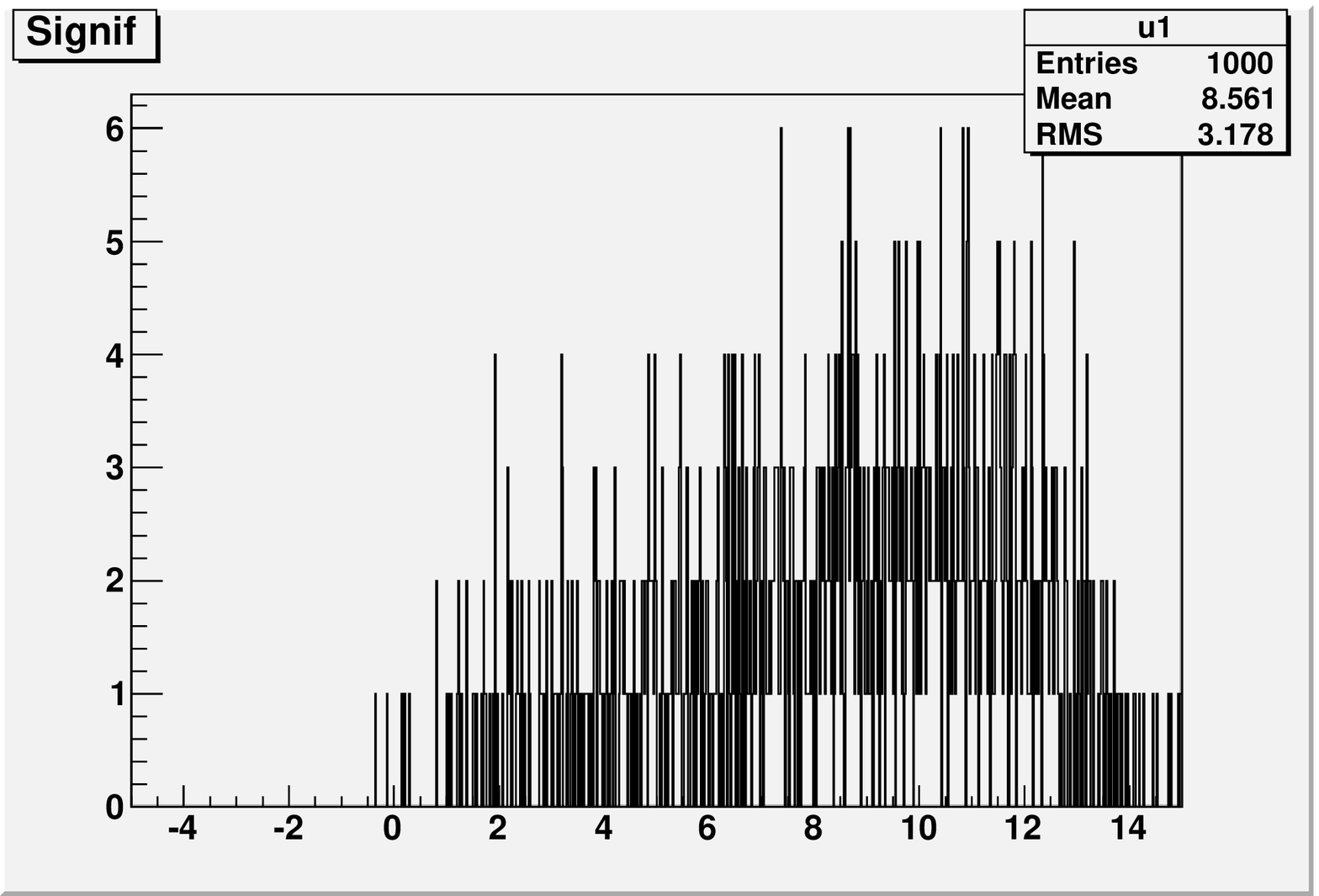}} 
\caption{Triangle distributions: the observed values $\hat n_{i1}$ in the first histogram (left,up), 
the observed values $n_{i2}$ in the second histogram (left, down), observed significances $\hat S_{i}$ 
bin-by-bin (right, up), the distribution of observed significances (right down).}
    \label{fig:4} 
\end{figure}

Suppose, the histograms are taken from experiments with different integrated luminosity, i.e. 
the total numbers of events in histograms are different. In this case the observed significances 
are  
changed from bin to bin (see, Fig.~\ref{fig:4}). Correspondingly, the distribution 
of significances has non-gaussian shape (in contrast with previous distributon of 
significances~(see, Fig.~\ref{fig:3})).

\begin{figure}[htpb]
  \begin{center}
           \resizebox{7.0cm}{!}{\includegraphics{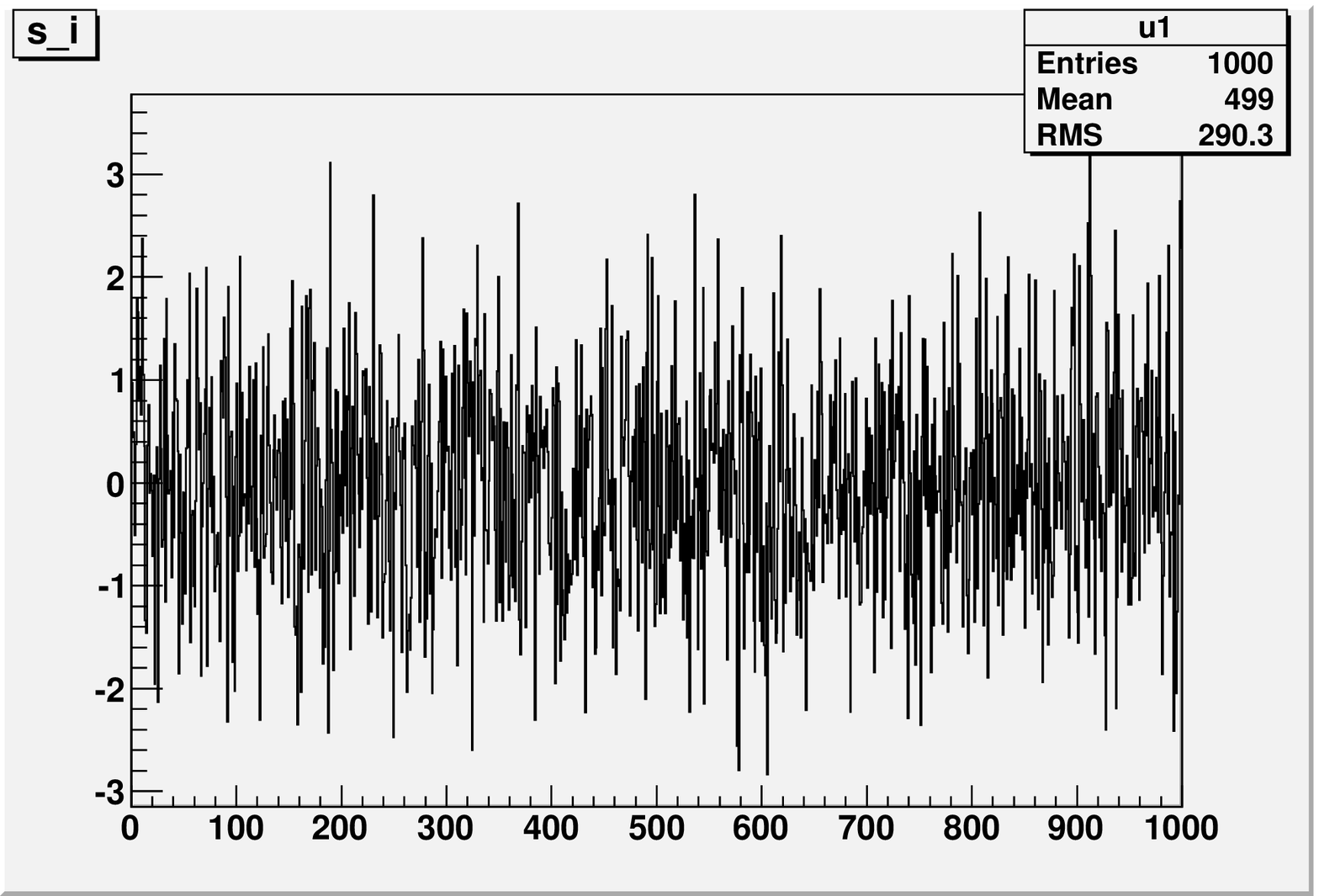}} 
           \resizebox{7.0cm}{!}{\includegraphics{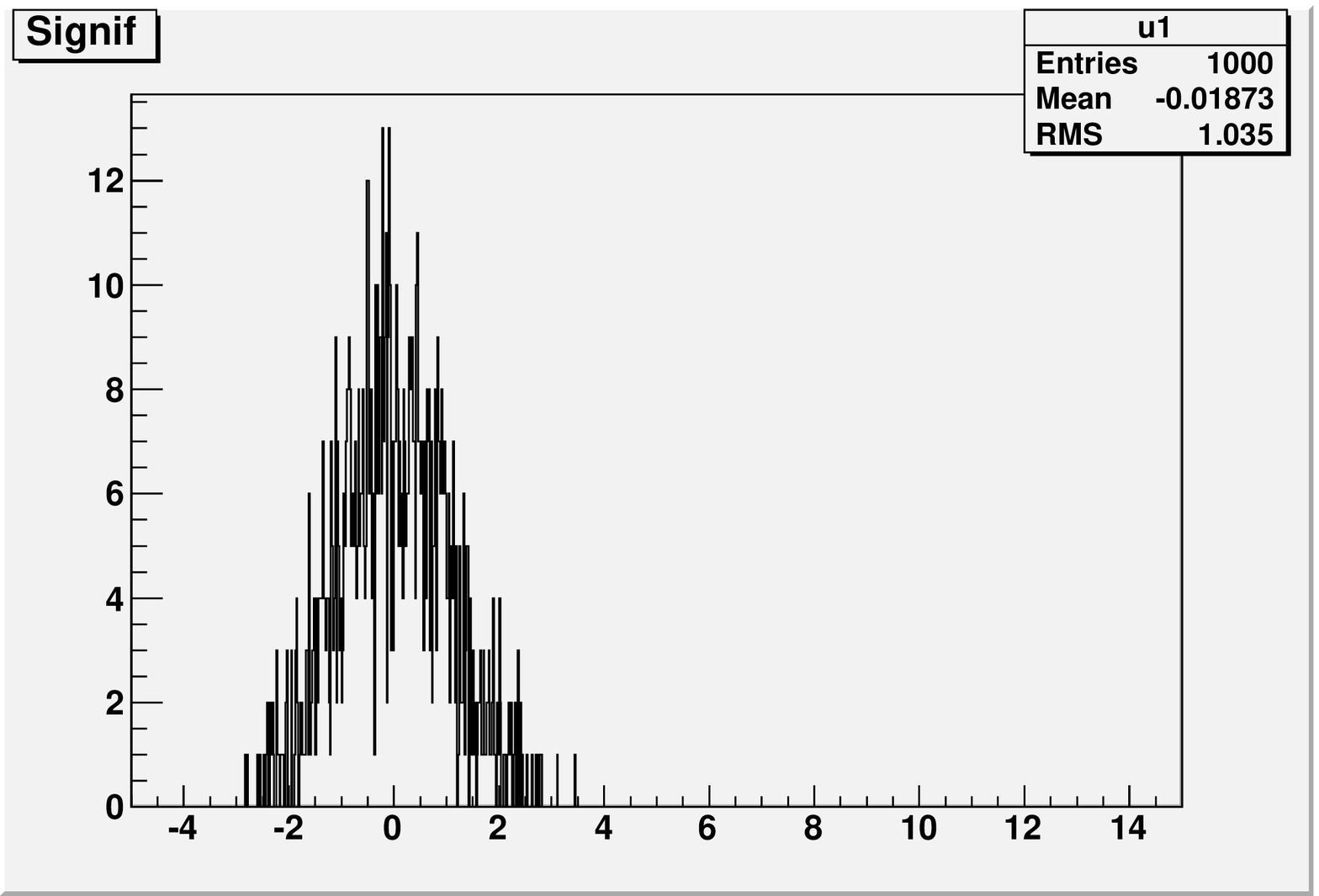}} 
\caption{Triangle distributions: observed normalized significances $\hat S_{i}$ bin-by-bin (left), 
the distribution of observed normalized significances (right).}
    \label{fig:5} 
  \end{center}
\end{figure}

For the normalized significance (Eq.~\ref{eq:3})  we have the standard 
normal distribution of significances (see, Fig.~\ref{fig:5}).

So, if two histograms are obtained from the same flow of events then the distribution 
of the normalized significances  obeys to the distribution which is close to the 
standard normal distribution. 
The $RMS$ of the distribution of significances is a measure of statistical difference between 
two histograms and, correspondingly, between two flows of events. This ``distance measure'' 
between two histograms has a clear interpretation: 

\begin{itemize}
\item $RMS=0$  -- histograms are identical;

\item $RMS\sim 1$  -- both histograms are obtained (by the using independent samples)
from the same parent distribution;

\item $RMS >> 1$ -- histograms are obtained from different parent distributions.

\end{itemize} 

Note, the relation~\footnote{Thanks to Luc Demortier.}
$$\displaystyle RMS^2 = \frac{\chi^2}{M} - \bar S^2,$$
where $\displaystyle \chi^2=\sum_{i=1}^M{\hat S_i^2}$,
exists for the distribution of significances. 

\section{Resolution of the method}

An accuracy (resolution) of the method depends on the number of bins $M$ in histograms, 
observed values in bins and, correspondingly, on the normalized coefficient $K$. 
The accuracy can be estimated via Monte Carlo experiments. 

Two models of the statistical populations (pseudo populations) are produced. 
Each of models represents one of the histograms. 
Namely, 4999 clones for each of histograms are produced 
by the Monte Carlo simulation of  content for each bin $i$ of histogram $k$  
due to the law $N(\hat n_{ik},\hat\sigma_{ik})$.      
It is possible because the normal distribution is a statistically 
dual distribution~\cite{StDual}.  
As a result there are 5000 pairs of histograms for comparisons. 

The comparison is performed for each pair of histograms 
(5000 comparisons). The distribution of the significances $\hat S_i$ 
($\hat S_i$ is a significance of deviation for bin $i$) is obtained as a result of  
each comparison. After that the $RMS$, $\bar S$ and 
$\sqrt{\frac{\chi^2}{M}}$~\footnote{It is done for comparison with $RMS$.}  
are calculated.

\begin{figure}
\includegraphics[width=0.5\textwidth]{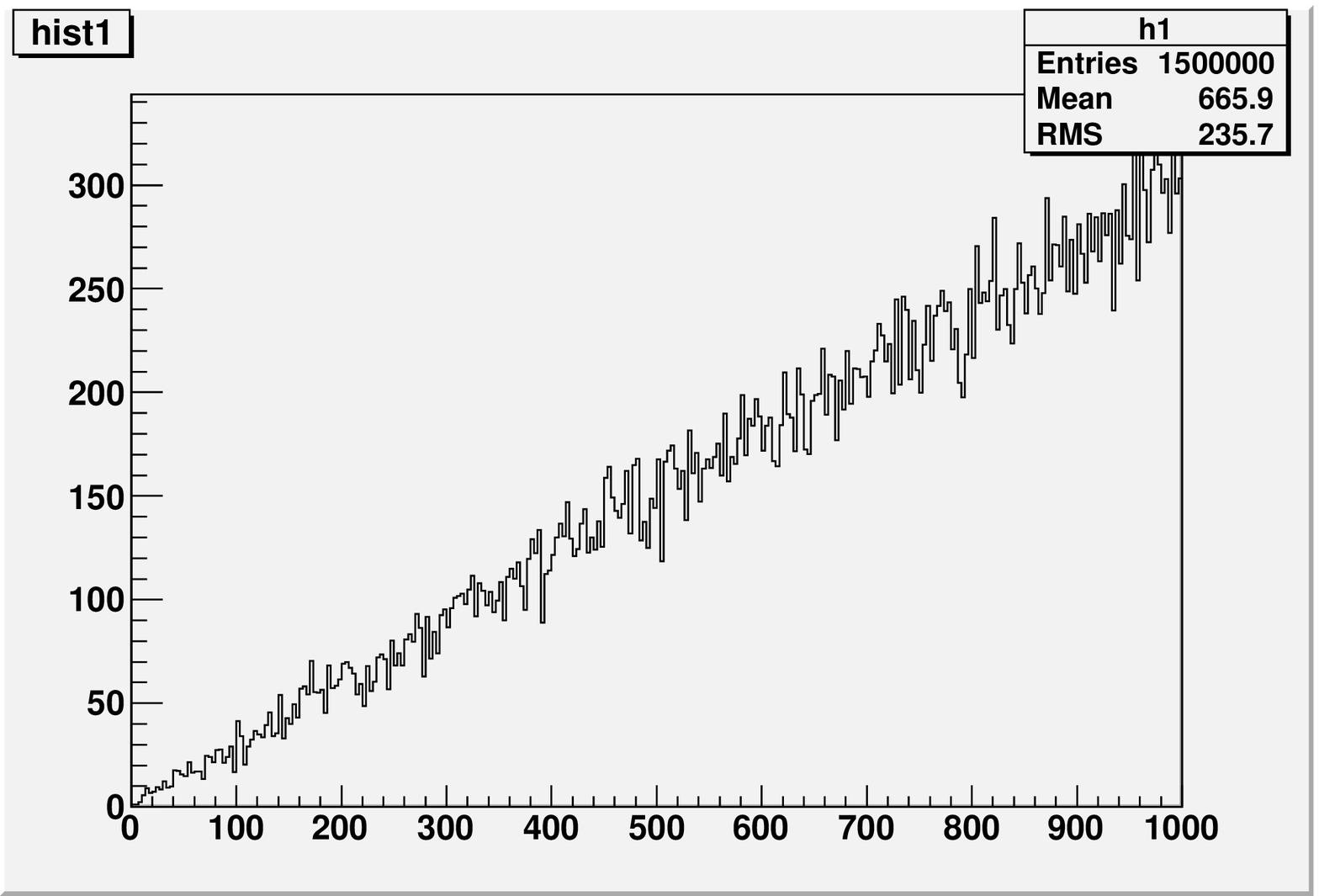} 
\includegraphics[width=0.5\textwidth]{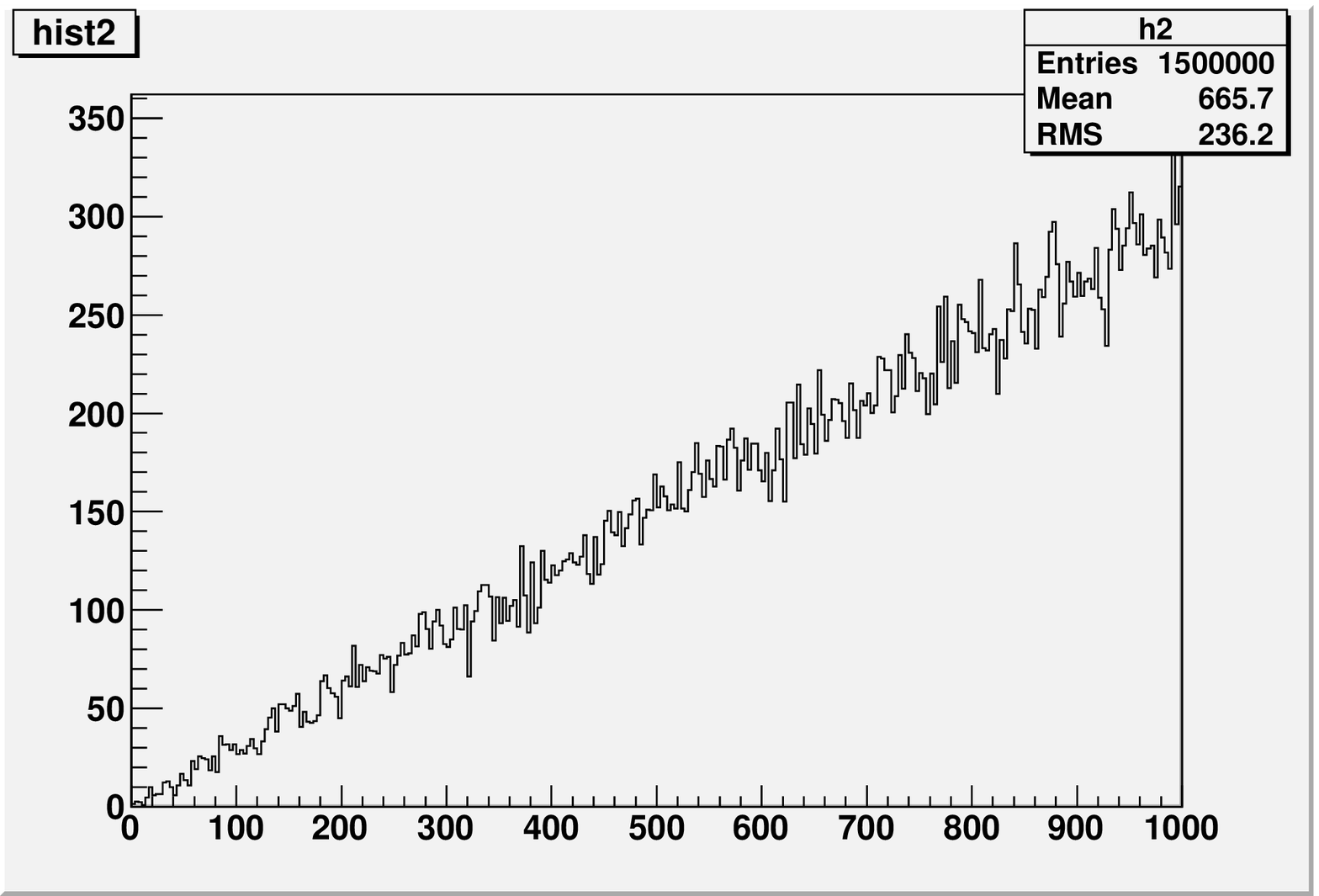} 
\caption{Case A: input histograms -- the same triangle distributions, M=300, K=1.}
    \label{fig:1Add} 
\end{figure}

\begin{figure}
\includegraphics[width=0.5\textwidth]{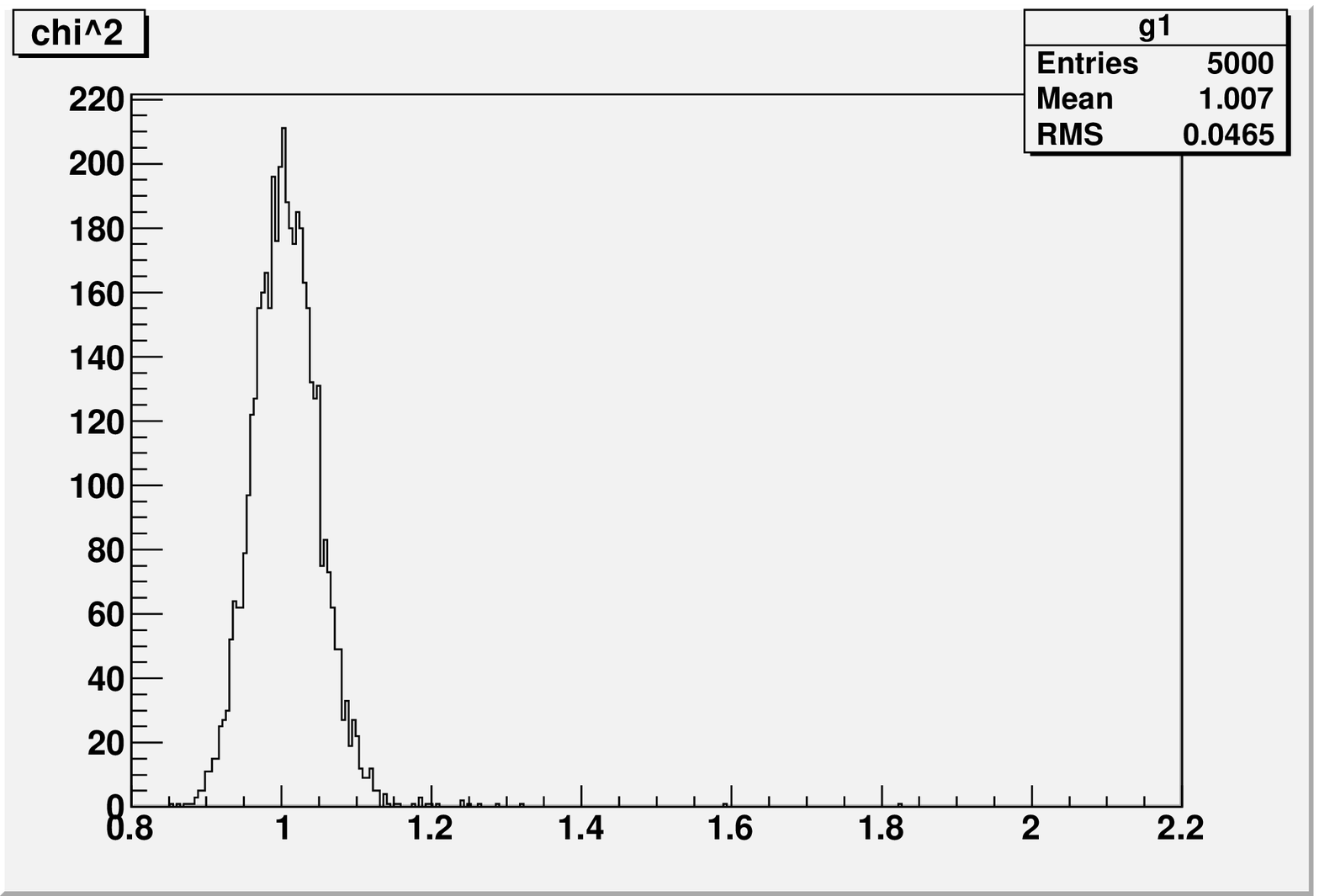} 
\includegraphics[width=0.5\textwidth]{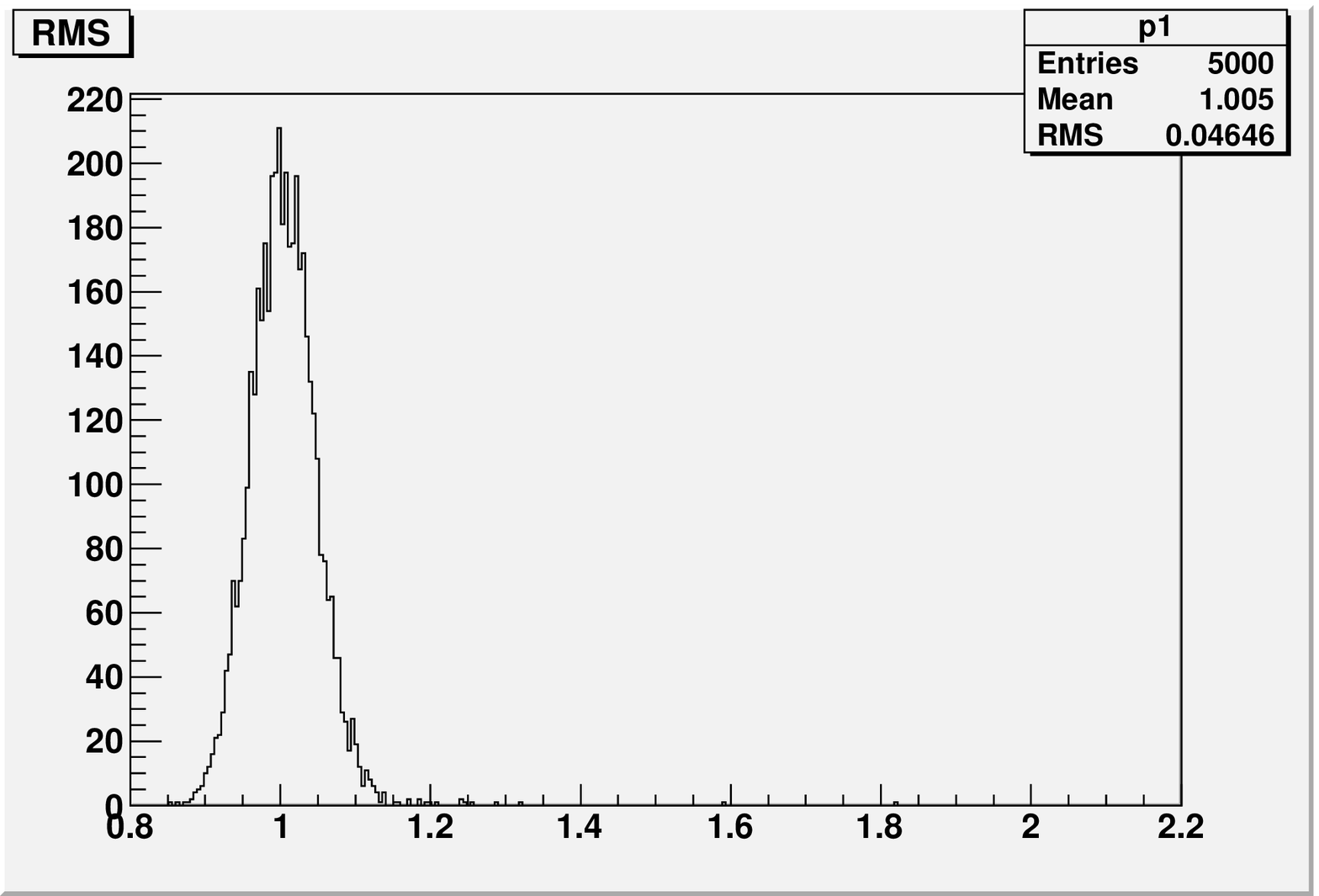} 
\caption{Case A: 5000 comparisons -- $\sqrt{\frac{\chi^2}{M}}$ for each trial (left), 
$RMS$ for each trial (right).}
    \label{fig:2Add} 
\end{figure}

\begin{figure}
\includegraphics[width=0.5\textwidth]{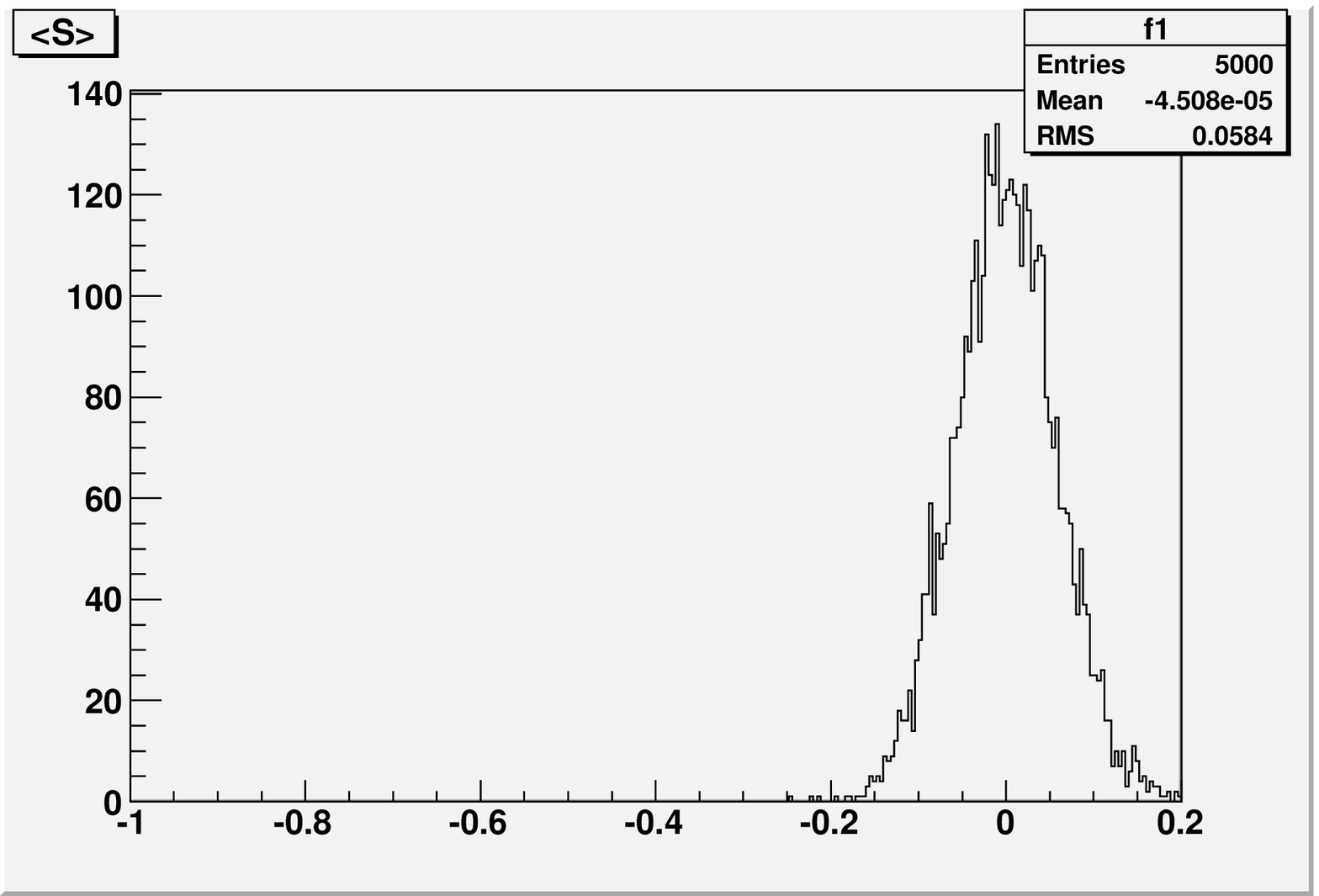} 
\includegraphics[width=0.5\textwidth]{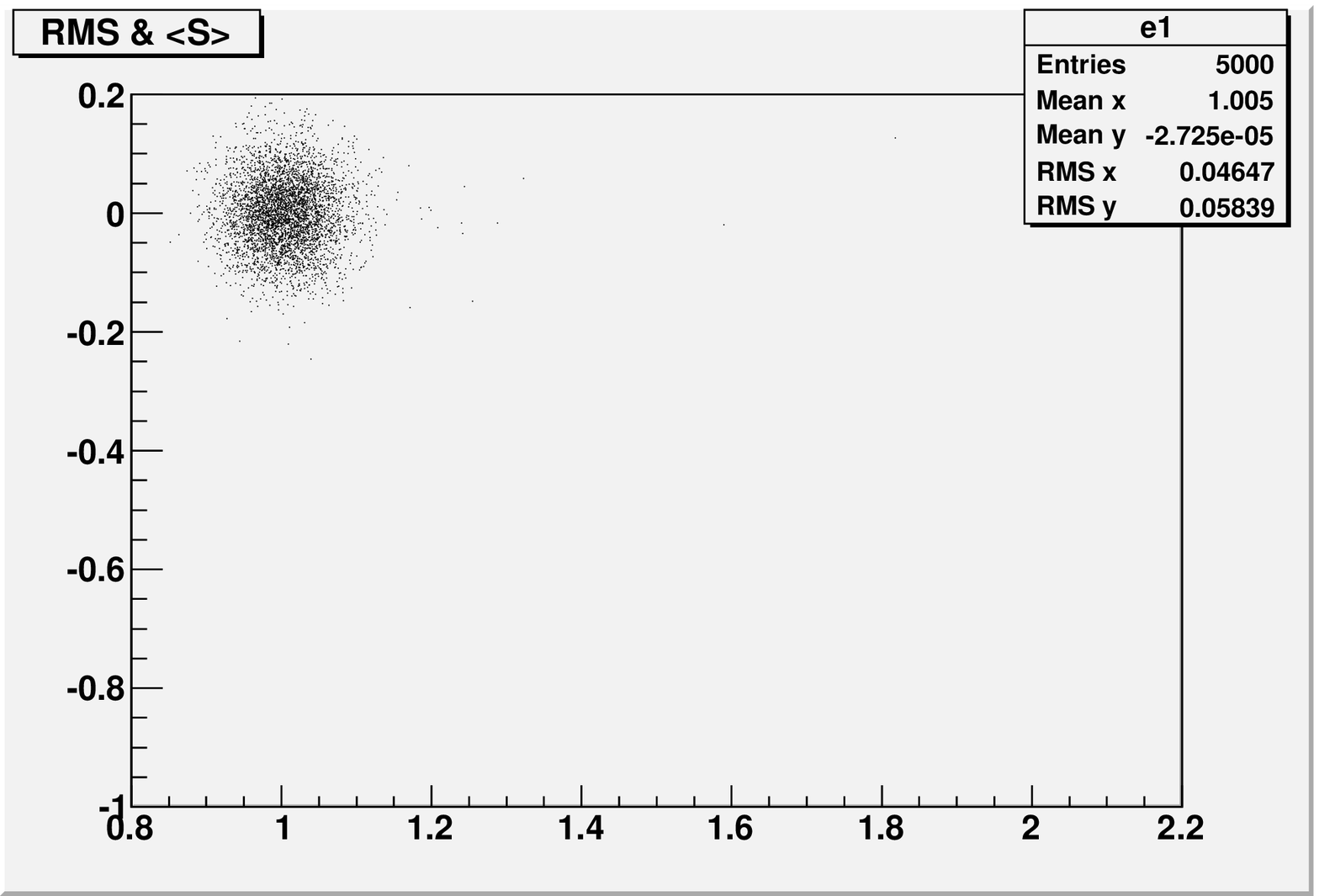} 
\caption{Case A: 5000 comparisons -- $\bar S$ for each trial (left), 
$RMS$ \& $\bar S$ (right).}
    \label{fig:3Add} 
\end{figure}

Let us consider the case (Case A) when both histograms are obtained from the same 
statistical population~(Fig.~\ref{fig:1Add}). The distributions of 
$\sqrt{\frac{\chi^2}{M}}$ (Fig.~\ref{fig:2Add}, left), $RMS$ (Fig.~\ref{fig:2Add}, right), 
$\bar S$ (Fig.~\ref{fig:3Add}, left) and $RMS$ versus $\bar S$ 
(Fig.~\ref{fig:3Add}, right) are produced during 5000 comparisons of histograms.

\begin{figure}[htbp]
\includegraphics[width=1.1\textwidth]{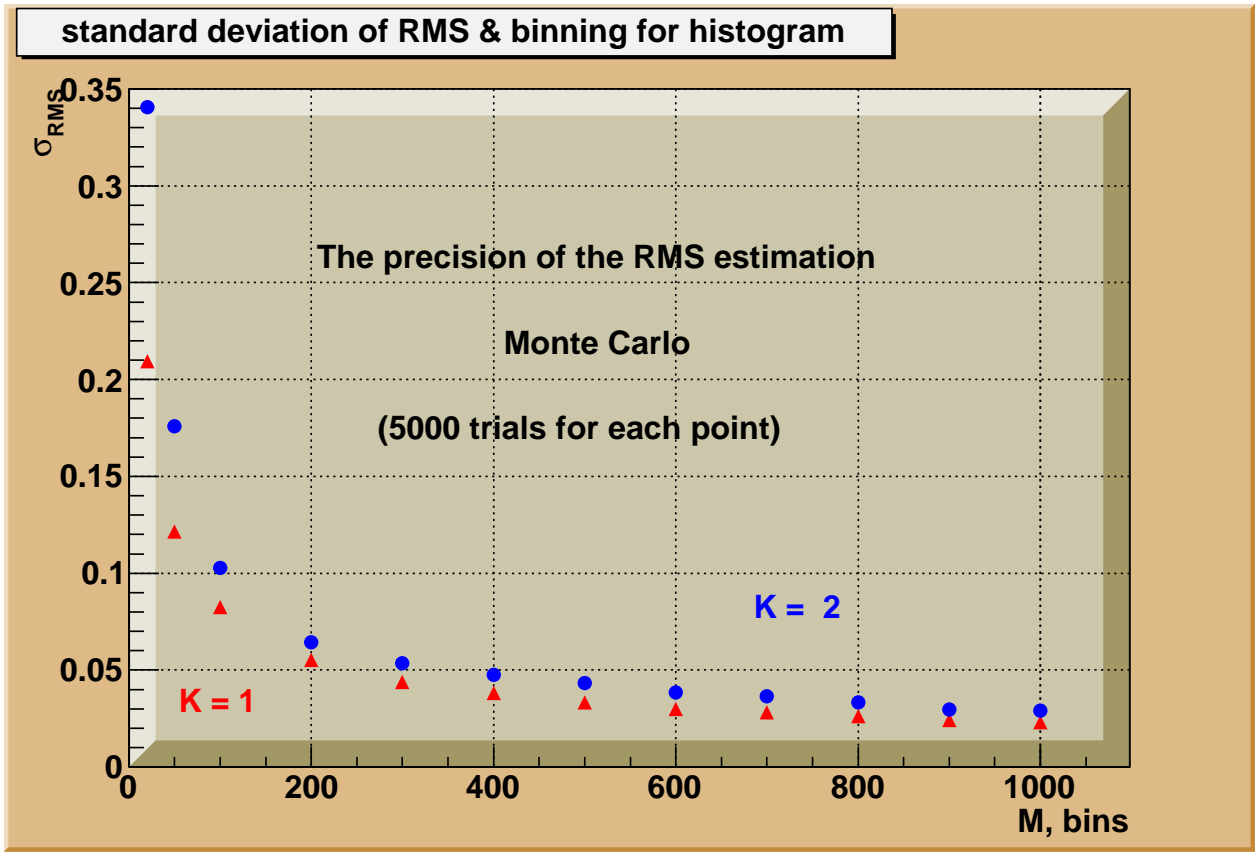} 
\caption{Triangle distributions (Sec.~4.2): the dependence of the standard 
deviation of RMS on number of bins $M$. This dependence is shown for two values of normalized 
coefficient $K$ ($K=1$ and $K=2$).}
    \label{fig:3ROOT} 
\end{figure}

The dependencies of the resolution ($RMS$) on the number of bins $M$ and on the 
value of coefficient of normalization $K$ are shown in Fig.~\ref{fig:3ROOT}. Note, 
the resolution of $\chi^2$ method~\cite{Porter} (Fig.~\ref{fig:2Add}, left) practically 
the same as resolution of $RMS$ in this method (Fig.~\ref{fig:2Add}, right).

\section{Distinguishability of histograms}

The estimation of the distinguishability of histograms is performed 
with the using of hypotheses testing.  
``A probability of correct decision'' about distinguishability of 
hypotheses~\cite{NIM534} is used as measure of the potential in separation  
of two histograms. 
It is probability of the correct choice between two hypotheses 
``the histograms are produced by the treatment of events from the same event 
flow (the same statistical population)'' 
or ``the histograms are produced by the treatment of events from different event flows''.  
This value can characterize the distinguishability of two histograms.
If $1 - \kappa = 1$ then the distinguishability of histograms  is 100\%, i.e. 
histograms are produced by the treatment of events from different event flows.    
If $1 - \kappa = 0$ then it is impossible to separate these histograms, i.e. 
histograms are produced by the treatment of events from the same event flow.    
The probability of correct decision $1 - \kappa$ is formed by the 
Type I error ($\alpha$) and by the Type II error ($\beta$) in hypotheses 
testing. $\alpha$ (Type I error) is the probability to accept the 
alternative hypothesis if the main hypothesis is correct. 
$\beta$ (Type II error) is the probability to accept the main 
hypothesis if the alternative hypothesis is correct. 
If critical region (critical value, critical line, ...) is used 
correctly, i.e. if $\alpha+\beta \le 1$, then 

$$1 - \displaystyle \kappa = 1 - \frac{\alpha+\beta}{2 - (\alpha+\beta)}.$$

\noindent

\begin{figure}
\includegraphics[width=0.5\textwidth]{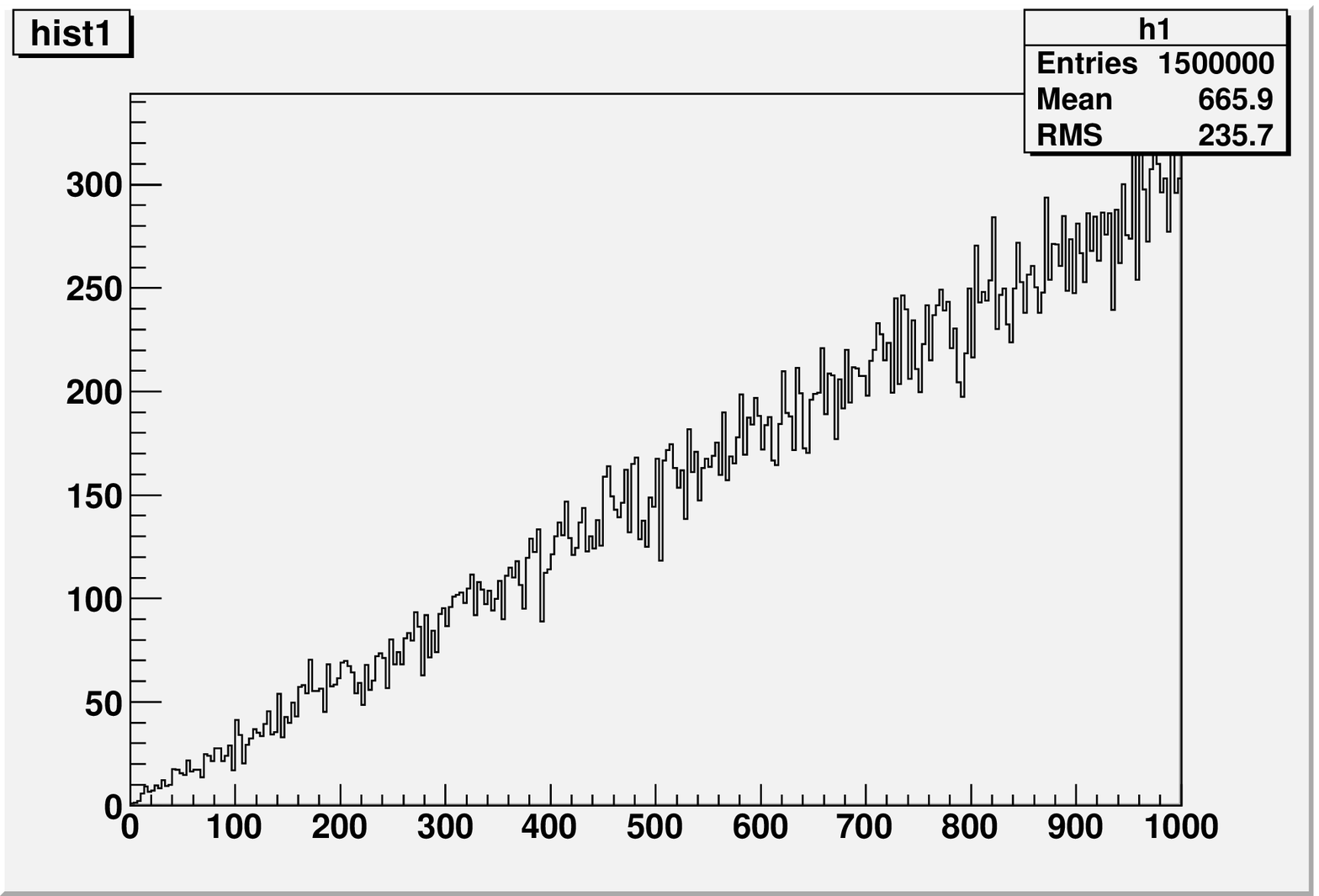} 
\includegraphics[width=0.5\textwidth]{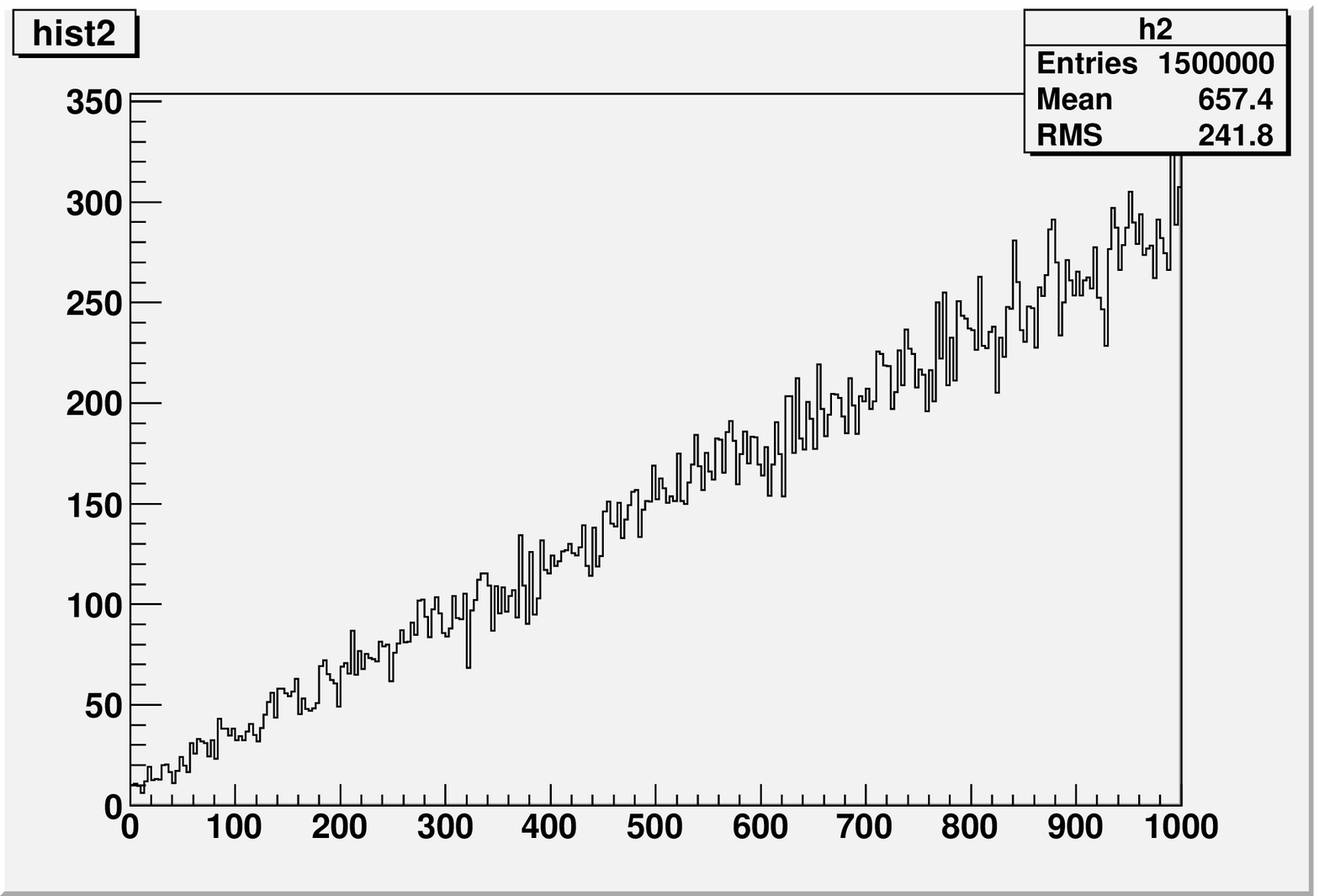} 
\caption{Case B: input histograms -- small difference between distributions, M=300, K=1.}
    \label{fig:4Add} 
\end{figure}

\begin{figure}
\includegraphics[width=0.5\textwidth]{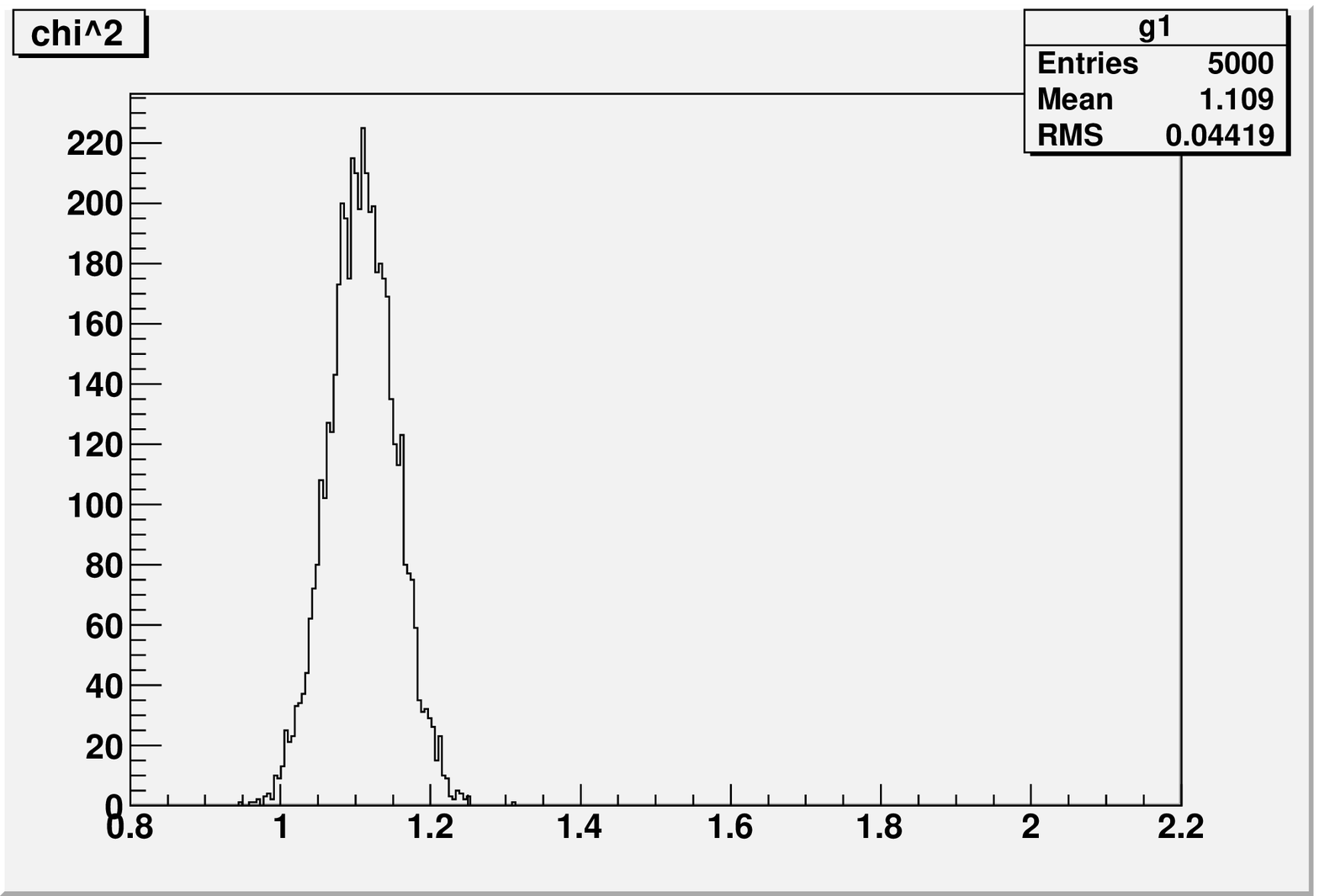} 
\includegraphics[width=0.5\textwidth]{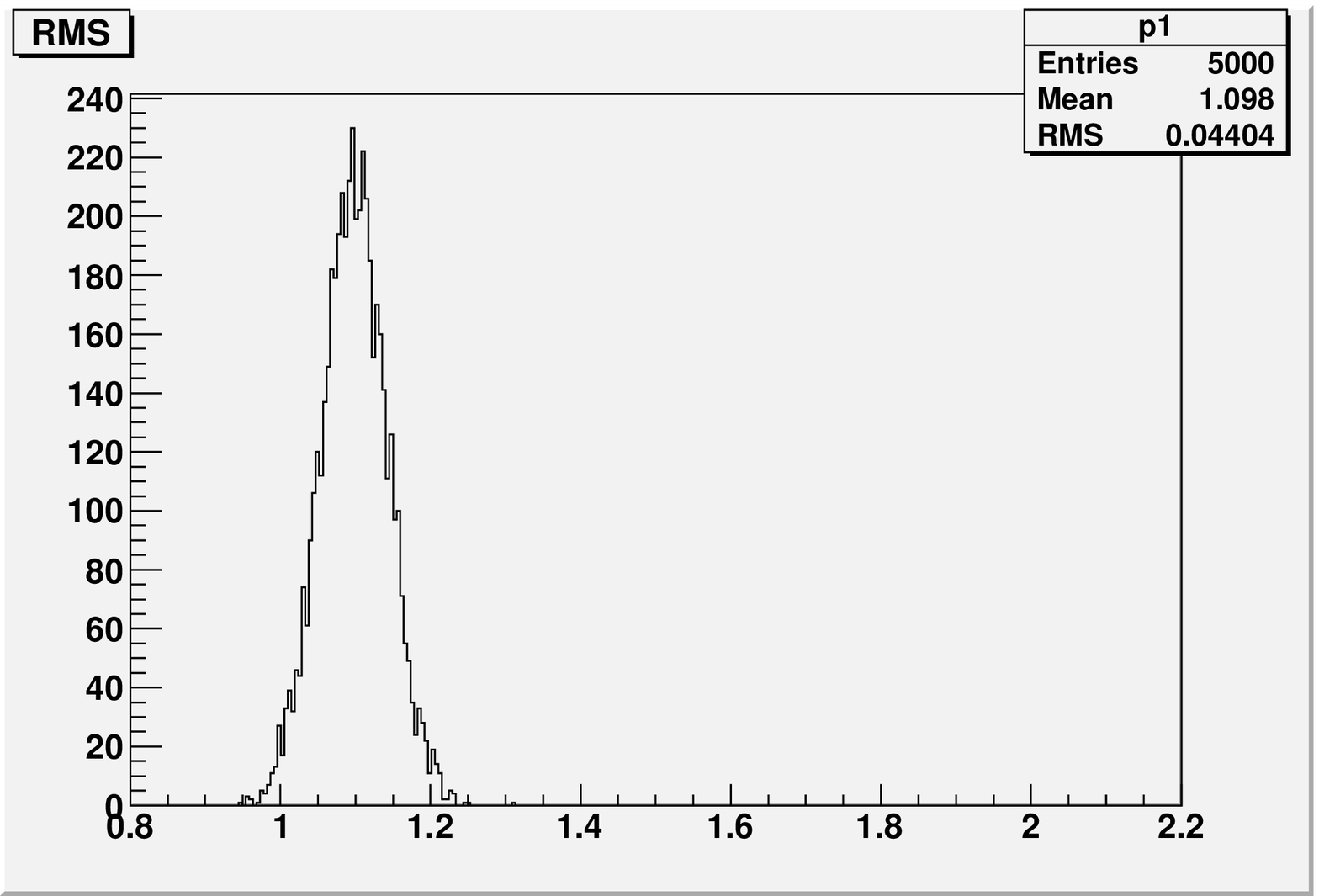} 
\caption{Case B: 5000 comparisons -- $\sqrt{\frac{\chi^2}{M}}$ 
for each trial (left), $RMS$ for each trial (right).}
    \label{fig:5Add} 
\end{figure}

\begin{figure}
\includegraphics[width=0.5\textwidth]{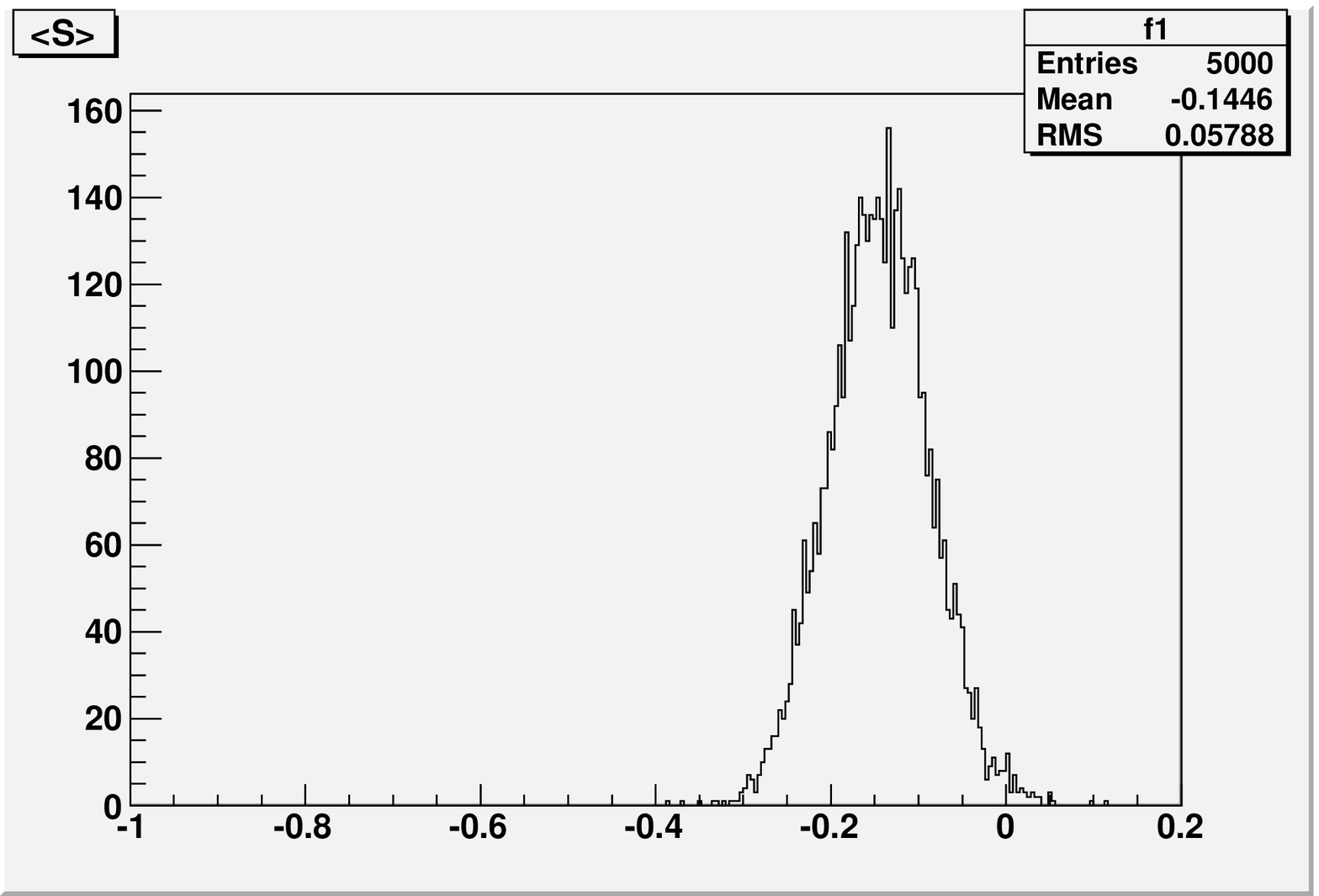} 
\includegraphics[width=0.5\textwidth]{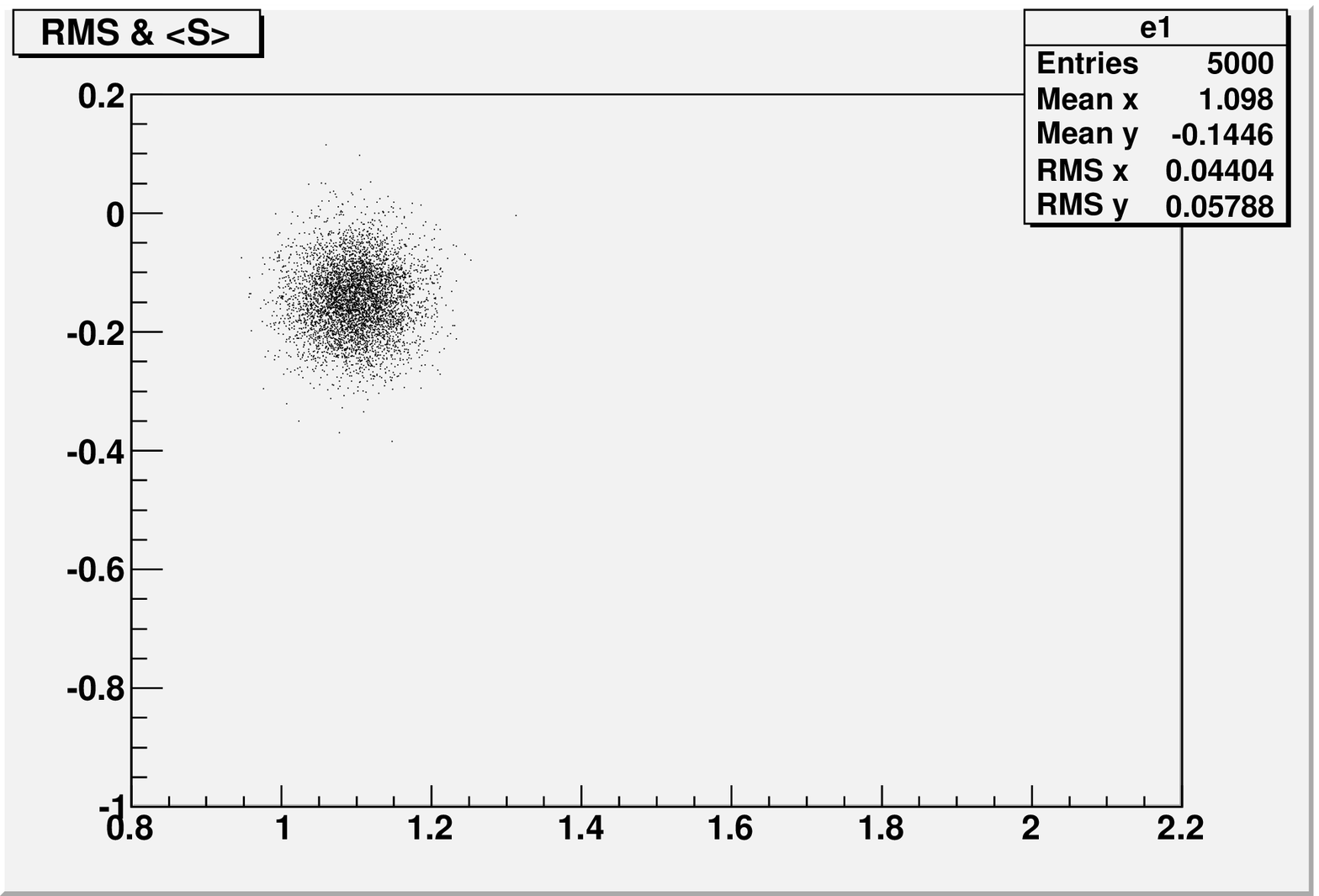} 
\caption{Case B: 5000 comparisons -- $\bar S$ for each trial (left), 
$RMS$ \& $\bar S$ (right).}
    \label{fig:6Add} 
\end{figure}

\begin{figure}
\includegraphics[width=0.5\textwidth]{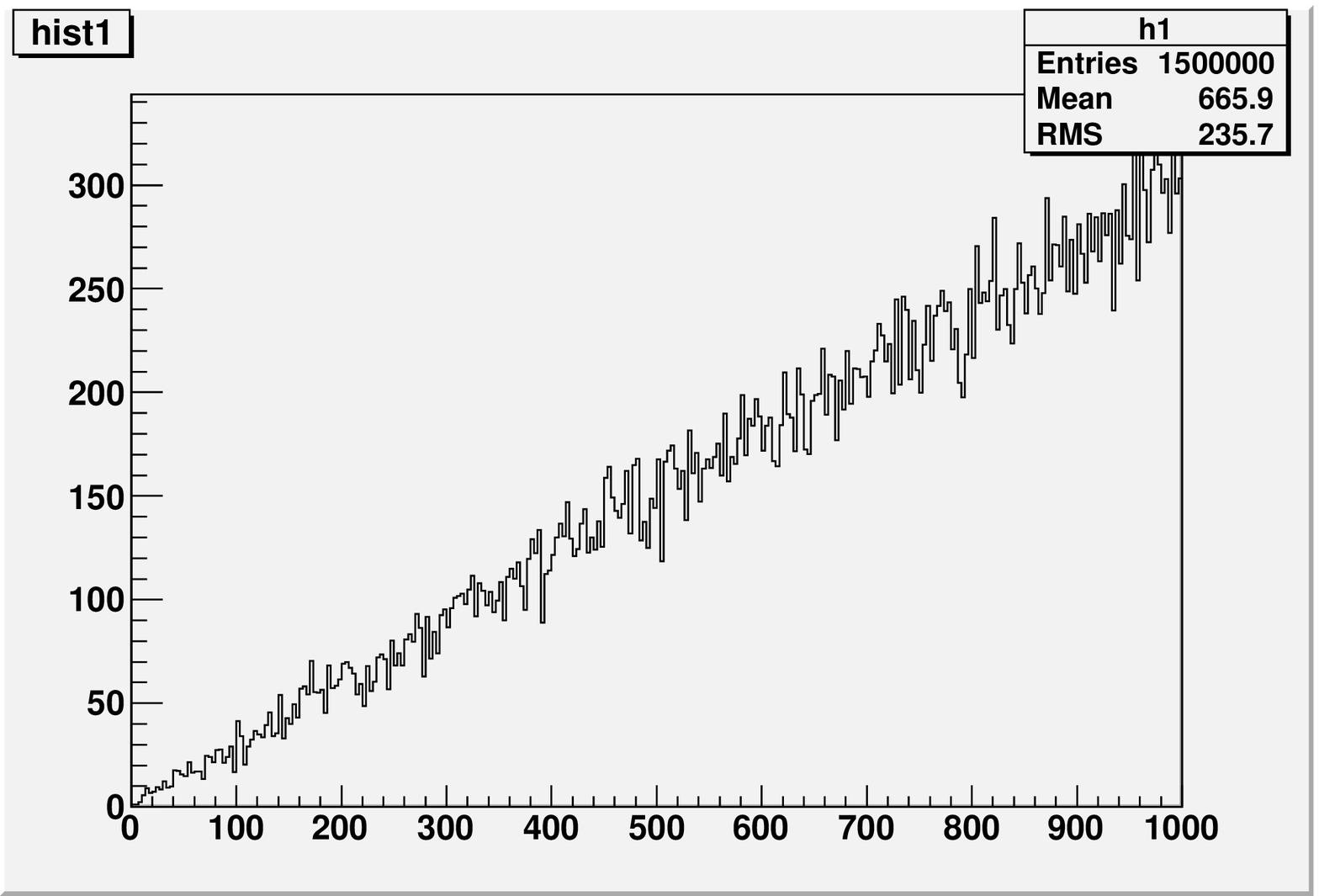} 
\includegraphics[width=0.5\textwidth]{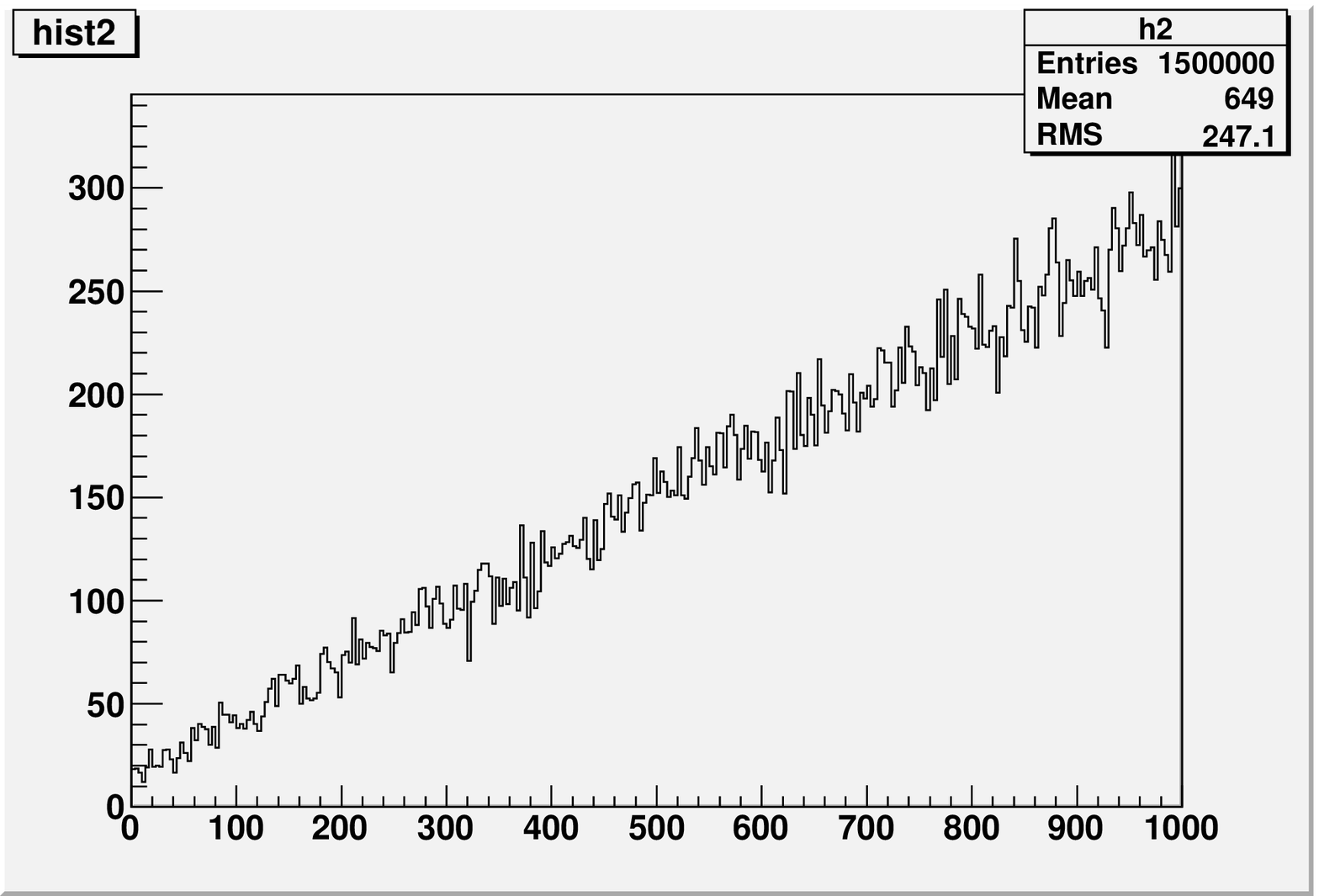} 
\caption{Case C: input histograms -- large difference between distributions, M=300, K=1.}
    \label{fig:7Add} 
\end{figure}

\begin{figure}
\includegraphics[width=0.5\textwidth]{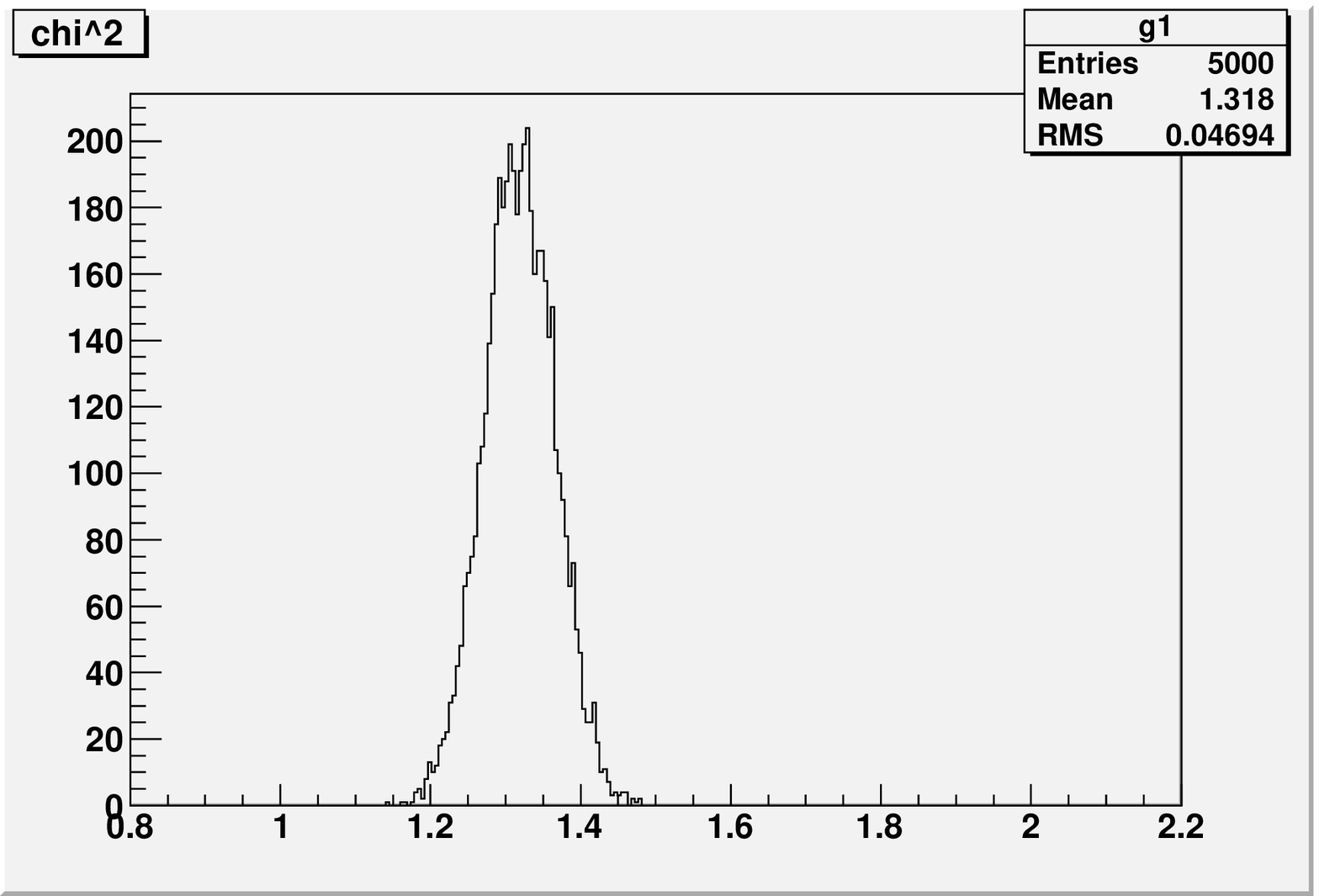} 
\includegraphics[width=0.5\textwidth]{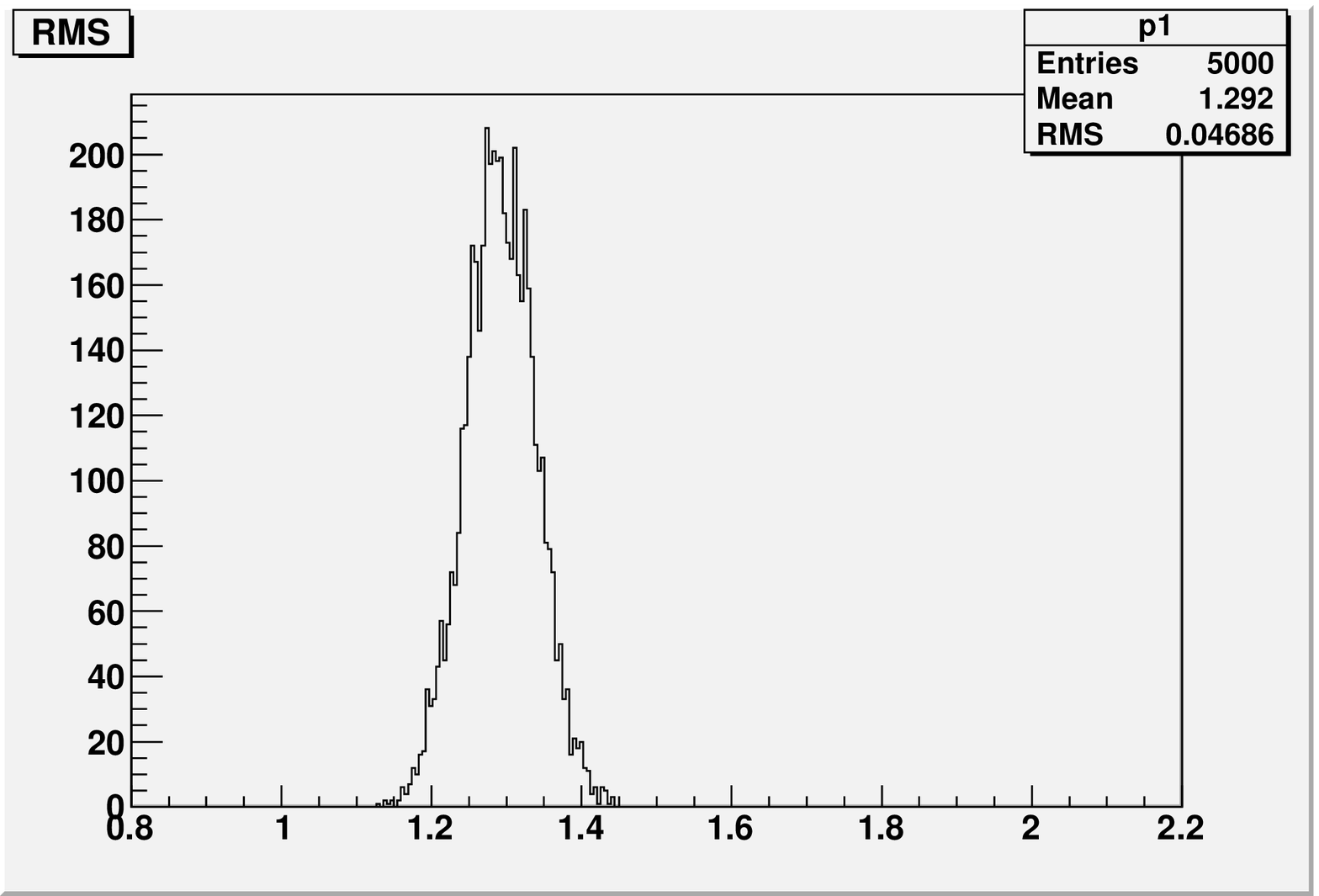} 
\caption{Case C: 5000 comparisons -- $\sqrt{\frac{\chi^2}{M}}$ for each trial (left), 
$RMS$ for each trial (right).}
    \label{fig:8Add} 
\end{figure}

\begin{figure}
\includegraphics[width=0.5\textwidth]{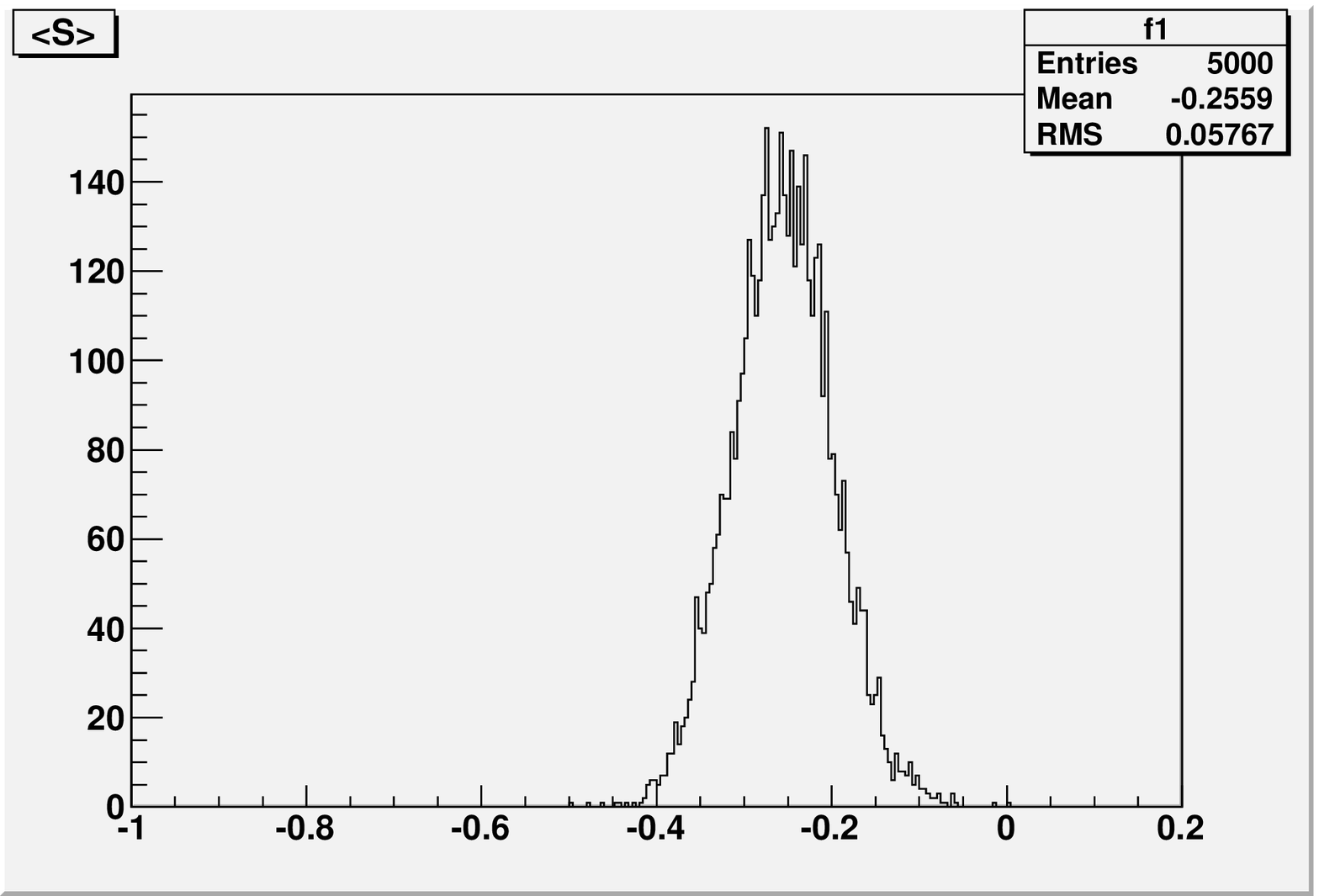} 
\includegraphics[width=0.5\textwidth]{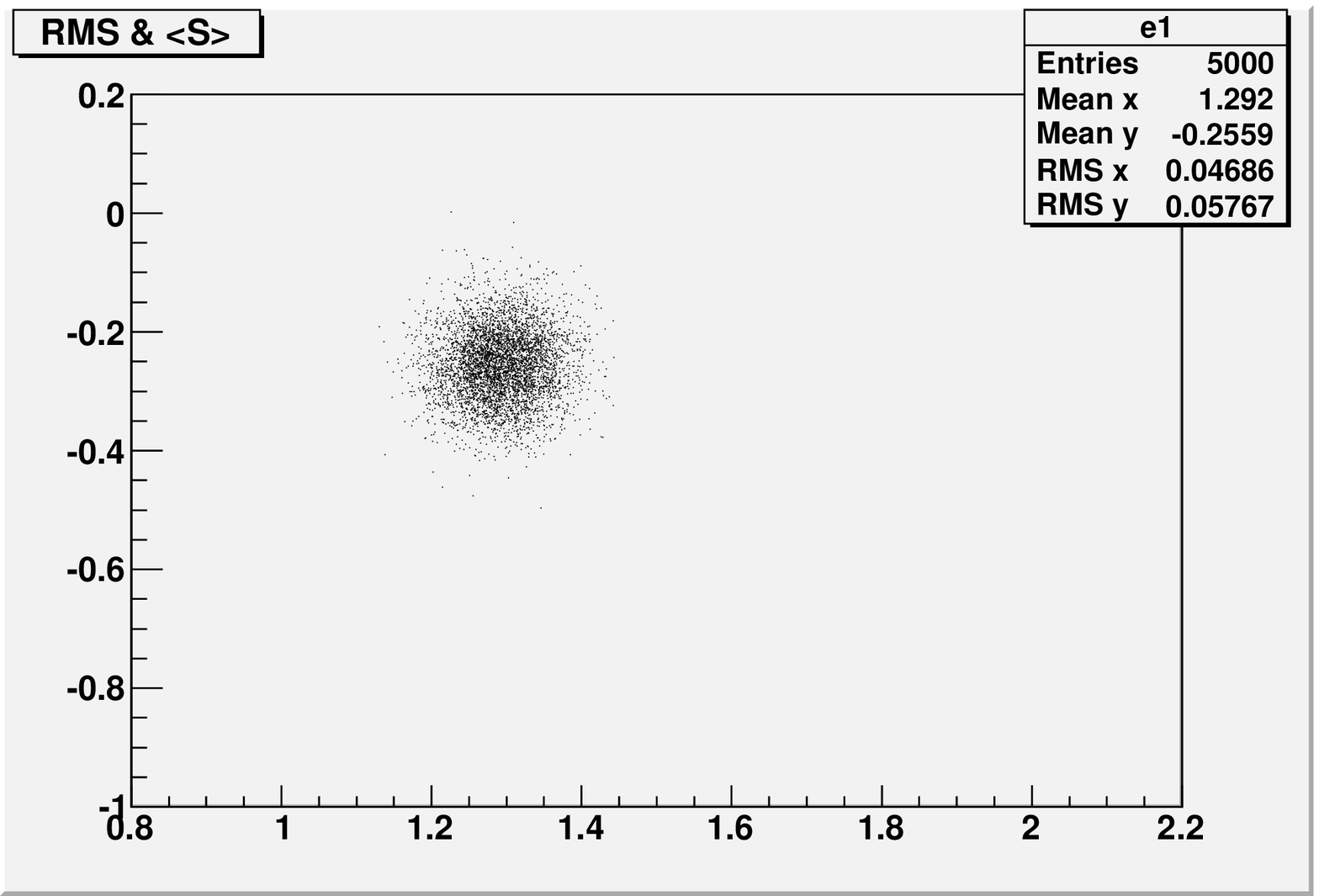} 
\caption{Case C: 5000 comparisons -- $\bar S$ for each trial (left), 
$RMS$ \& $\bar S$ (right).}
    \label{fig:9Add} 
\end{figure}

Two additional cases are served to illustrate the 
estimation of distinguishability of histograms.  
The Case B (Fig.~\ref{fig:4Add}): the first histogram has a triangle 
distribution, the slope of the distribution in the second histogram  
is changed a bit with the same integral under distribution. 
The distributions of $\sqrt{\frac{\chi^2}{M}}$ (Fig.~\ref{fig:5Add}, left), 
$RMS$ (Fig.~\ref{fig:5Add}, right), $\bar S$ (Fig.~\ref{fig:6Add}, left) and 
$RMS$ versus $\bar S$ (Fig.~\ref{fig:6Add}, right) also 
are produced during 5000 comparisons of histograms.
The Case C (Fig.~\ref{fig:7Add}): the first histogram has a 
triangle distribution, the slope of the distribution in the second 
histogram  is changed more significantly than in the case B 
(integral is the same). Correspondingly, the distributions of 
$\sqrt{\frac{\chi^2}{M}}$ (Fig.~\ref{fig:8Add}, left), 
$RMS$ (Fig.~\ref{fig:8Add}, right), $\bar S$ (Fig.~\ref{fig:9Add}, left) 
and $RMS$ versus $\bar S$ (Fig.~\ref{fig:9Add}, right)  
are produced during 5000 comparisons of histograms.

\newpage 

The probability of the correct decision as a measure for 
distinguishability of two histograms is determined by the 
comparison of distribution for the Case A and corresponding 
distribution for the Case B or Case C. The critical value is used 
for comparison of  one-dimensional $\sqrt{\frac{\chi^2}{M}}$ distributions. 
The critical line is used for comparison of two-dimensional 
$RMS\&\bar S$ distributions. The results are presented in 
Tab.~\ref{tab:1}, Tab.~\ref{tab:2}, Tab.~\ref{tab:3} 
and  Tab.~\ref{tab:4}.

For $\chi^2$ method the probability of the correct decision 
($1-\kappa$)  about the Case realization (A or B) equals 87.26\%.
For another method the probability of the correct decision 
($1-\kappa$)  about the Case realization (A or B) equals 93.88\%.
One can see that the method, which uses $RMS$ and $\bar S$, 
gives better distinguishability of histograms than the $\chi^2$ method. 
Note, in this study are used only two moments of the distribution of significances 
(the first initial moment ($\bar S$) and the root from the second central 
moment ($RMS$)) for estimation of distinguishability of histograms. 

\section{Conclusions}

The proposed approach allows to perform the comparison of histograms 
in more detail than methods which use only one test statistics. 
This method can be used in tasks of monitoring of the equipment 
during experiments. The first experience of the using the method 
is presented in report~\cite{ROOT2013}.  
Note, the production of statistical (pseudo) populations, which represent 
the comparing histograms, allows to estimate the distinguishability 
of histograms in each comparison.

\acknowledgments

The authors are grateful to L. Demortier, T. Dorigo, L.V.~Dudko, V.A.~Kachanov, 
L. Lyons, V.A.~Matveev, L.~Moneta and E. Offermann for the interest and useful comments. 
The authors would like to thank Yu. Gouz, E. Gushchin, A. Karavdina, 
D.~Konstantinov, A.~Popov and N.~Tsirova for fruitful discussions. 
This work is supported by RFBR grant N 13-02-00363.

\begin{table}[htb]
\begin{center}
\begin{tabular}{r|rrr} 
\hline
\hline
         &        & In      & reality \\  
Accepted & Case A &  Case B & \\
\hline
Case A   &   4543 &  673    &  \\
Case B   &    457 & 4327    &  \\
\hline
\hline
 $1-\kappa$ & $\alpha$ & $\beta$ & \\
\hline
0.8726      & 0.0914   &  0.1346 &  \\
\hline
\end{tabular}
\caption{ $\sqrt{\frac{\chi^2}{M}}$ - 5000 decisions.}
\label{tab:1}
\end{center}
\end{table}

\begin{table}[htb]
\begin{center}
\begin{tabular}{r|rrr}
\hline
\hline
         &        &  In      & reality \\  
Accepted & Case A &  Case B  & \\
\hline
Case A   &   4843 &  456     & \\
Case B   &    121 & 4544     & \\
\hline
\hline
$1-\kappa$ & $\alpha$ & $\beta$ & \\
\hline
0.9388     & 0.0242   &  0.0912 & \\
\hline
\end{tabular}
\caption{ $RMS\&\bar S$ - 5000 decisions.}
\label{tab:2}
\end{center}
\end{table}

\begin{table}[htb]
\begin{center}
\begin{tabular}{r|rrr}
\hline
\hline
         &        &  In      & reality \\  
Accepted & Case A &  Case C  & \\
\hline
Case A   &   4982 &    1     &  \\
Case C   &     18 & 4999    &  \\
\hline
\hline
$1-\kappa$ & $\alpha$ & $\beta$ & \\
\hline
0.9981     & 0.0036   &  0.0002 & \\
\hline
\end{tabular}
\caption{ $\sqrt{\frac{\chi^2}{M}}$ - 5000 decisions.}
\label{tab:3} 
\end{center}
\end{table}

\begin{table}[htb]
\begin{center}
\begin{tabular}{r|rrr}
\hline
\hline
         &        & In       & reality \\  
Accepted & Case A &  Case C  & \\
\hline
Case A   &   4989 &    0     & \\
Case C   &     11 & 5000     & \\
\hline
\hline
$1-\kappa$ & $\alpha$ & $\beta$ & \\
\hline
0.9989     & 0.0022   &  0.0000 &  \\
\hline
\end{tabular}
\caption{ $RMS\&\bar S$ - 5000 decisions.}
\label{tab:4}
\end{center}
\end{table}


\begin{thebibliography}{99}


\bibitem{Porter} F. Porter, {\it Testing Consistency of Two Histograms}, arXiv:0804.0380.


\bibitem{Gagun} N.D.~Gagunashvili, {\it Chi-square tests for comparing weighted histograms}, 
\emph{Nucl.Instr.\&Meth.}, {\bf A614} (2010) 287-296; arXiv:0905.4221.   


\bibitem{ACAT08}  S.I.~Bityukov, N.V.~Krasnikov, A.N.~Nikitenko, V.V.~Smirnova.   
{\it Two approaches to Combining Significances}. 
\emph{Proceedings of Science}, PoS ({\bf ACAT08}) 118.

\bibitem{Frod}  A.G. Frodesen, O. Skjeggestad, H. Toft, 
{\it Probability and Statistics in Particle Physics}, 
UNIVERSITETSFORLAGET, Bergen-Oslo-Troms$\o$, 1979. 

\bibitem{ROOT}  R.~Brun, F.~Rademaker, 
{\it ROOT -- An object oriented data analysis framework}, 
\emph{Nucl.Instr.\&Meth.}, {\bf A389} (1997) 81-86.

\bibitem{StDual} S.I.Bityukov, N.V. Krasnikov, V.A. Taperechkina, 
V.V. Smirnova, \emph{Statistically dual distributions in statistical 
inference}, in proceedings of \emph{Statistical problems in Particle 
Physics, Astrophysics and Cosmology} (PhyStat'05), September 12-15, 
2005, Oxford, UK, Imperial College Press, 2006, pp.102-105.

\bibitem{NIM534} 
S.I.~Bityukov, N.V.~Krasnikov,    
{\it Distinguishability of Hypotheses},  
Nucl.Inst.\&Meth. {\bf A534} (2004) 152-155.

\bibitem{ROOT2013} 
S.~Bityukov, N.~Tsirova, 
{\it Program for statistical comparison of histograms}, 
ROOT Users Workshop, 11 - 14 March, Saas-Fee, Switzerland,
https://indico.cern.ch/contributionDisplay.py?contribId=35\&confId=217511
     
\end{thebibliography}
\end{document}